\documentclass[12pt]{article}

\usepackage[left=2cm,right=2cm,top=2.8cm,bottom=2.8cm]{geometry}

\usepackage[colorlinks=true,linkcolor=black,citecolor=black,urlcolor=blue,filecolor=black]{hyperref}

\usepackage{tensor}
\usepackage{amsmath}
\usepackage{amsfonts}
\usepackage{amssymb,amsthm,mathabx}
\usepackage{color}
\usepackage{ytableau}
\usepackage{enumerate}
\usepackage{cite}

\numberwithin{equation}{section}

\def\a{\alpha}

\def\d{\delta}

\def\h{\eta}

\def\Th{\Theta}

\newcommand{\half}{\frac{1}{2}}
\renewcommand{\d}{\partial}

\newcommand{\ffrac}[2]{\raisebox{.5pt}
  {\footnotesize$\displaystyle\frac{#1}{#2}$}\kern1pt}

\newcommand{\ddl}[2]{\ffrac{\d #1}{\d #2}}
\newcommand{\ddll}[2]{\ffrac{\d^L #1}{\d #2}}

\newcommand{\vddr}[2]{\ffrac{\delta^R #1}{\delta #2}}
\newcommand{\vddl}[2]{{\ffrac{\delta #1}{\delta #2}}}

\def\cA{{\cal A}}

\def\cI{{\cal I}}

\def\cL{{\cal L}}

\def\cR{{\cal R}}

\def\cV{{\cal V}}

\def\be{\begin{equation}}
\def\ee{\end{equation}}
\def\bea{\begin{eqnarray}}
\def\eea{\end{eqnarray}}
\def\ba{\begin{array}}
\def\ea{\end{array}}

\def\nn{\nonumber}

\newcommand{\R}{\mathbb{R}}

\def\12{\frac{1}{2}}

\ytableausetup{mathmode, boxsize=1.3em}

\begin{document}

\vspace{30pt}

\begin{center}

{\Large\sc Deformations of vector-scalar models} 

\vspace{25pt} {\sc Glenn Barnich${}^{\, a}$, Nicolas
  Boulanger${}^{\, a, b}$, Marc Henneaux${}^{\, a, c}$,\\ 
  Bernard Julia${}^{\, c}$, Victor Lekeu${}^{\, a}$ and Arash
  Ranjbar${}^{\, a}$}

\vspace{10pt} {${}^a$\sl\small Universit\'e libre de Bruxelles and
  International Solvay Institutes, Campus Plaine CP231, B-1050
  Brussels, Belgium \vspace{10pt}

  ${}^b$\sl \small Groupe de M\'ecanique et Gravitation, Physique
  Th\'eorique et Math\'ematique,
  Universit\'e de Mons -- UMONS, 20 Place du Parc, B-7000 Mons, Belgium\\
  \vspace{10pt}

  ${}^c$\sl \small Laboratoire de Physique th\'eorique de l'Ecole
  Normale Sup\'erieure, 24 rue Lhomond, \\ 
  75231 Paris CEDEX, France\\
  \vspace{10pt} }

\vspace{40pt} {\sc\large Abstract} 
\end{center}

\noindent

Abelian vector fields non-minimally coupled to uncharged scalar fields
arise in many contexts.  We investigate here through algebraic methods
their consistent deformations (``gaugings''), i.e., the deformations
that preserve the number (but not necessarily the form or the algebra)
of the gauge symmetries. Infinitesimal consistent deformations are
given by the BRST cohomology classes at ghost number zero.  We
parametrize explicitly these classes in terms of various types of
global symmetries and corresponding Noether currents through the
characteristic cohomology related to antifields and equations of
motion. The analysis applies to all ghost numbers and not just ghost
number zero. We also provide a systematic discussion of the linear and
quadratic constraints on these parameters that follow from
higher-order consistency. Our work is relevant to the gaugings of
extended supergravities.

\newpage

\tableofcontents

\newpage

\section{Introduction}\label{sec:intro}

Our paper is devoted to a systematic study of the consistent
deformations of the gauge invariant actions of the form
\begin{equation}
  \label{eq:1bis}
  S_0[A_\mu^I,\phi^i]=\int \!d^4\!x\, \mathcal{L}_0,
\end{equation}
depending on $n_s$ uncharged scalar fields $\phi^i$ and $n_v$ abelian vector
fields $A_\mu^I$.  We assume that the only gauge symmetries of
(\ref{eq:1bis}) are the standard $U(1)$ gauge transformations for each
vector field, so that the gauge algebra is abelian and given by $n_v$
copies of $\mathfrak{u}(1)$.  A generating set of gauge invariances
can be taken to be
\begin{equation}
  \label{eq:2bis}
  \delta A_\mu^I=\d_\mu\epsilon^I,\quad \delta\phi^i =0. 
\end{equation}
The Lagrangian takes the form
\begin{equation}
  \mathcal{L}_0= \mathcal{L}_S[\phi^i]+\mathcal{L}_V[A^I_\mu,\phi^i],  \label{eq:Starting}
\end{equation}
where $ \mathcal{L}_V$ is a function that depends on the vector fields
through the abelian curvatures
$F^I_{\mu\nu}=\d_\mu A_\nu^I-\d_\nu A^I_\mu$ only, and which can also
involve the scalar fields $\phi^i$.  Derivatives of these variables
are in principle allowed in the general analysis carried out below,
but actually do not occur in the explicit Lagrangians discussed in
more detail.  The scalar fields can occur non linearly, e.g. terms of
the form $\mathcal{I}_{IJ}(\phi) F^I_{\mu\nu} F^{J\mu\nu}$ where
$\mathcal{I}_{IJ}(\phi)$ are some functions of the $\phi^i$'s are
allowed.  Similarly, the scalar Lagrangian need not be quadratic.
More on this in Subsection \ref{sec:model}.

The gauge transformations (\ref{eq:2bis}) are sometimes called ``free abelian gauge transformations'' to emphasize that the scalar fields are uncharged and do not transform under them.  This does not mean that the abelian vector fields themselves are free since non linear terms (non minimal couplings) are allowed in (\ref{eq:Starting}).

This class of models contains the vector-scalar sectors of
``ungauged'' extended supergravities, of which ${\mathcal N}=4$
\cite{Das:1977uy,Cremmer:1977tc,Cremmer:1977tt} and ${\mathcal N}=8$
\cite{Cremmer:1978km,Cremmer:1979up} supergravities offer prime
examples. These will be considered in detail in Sections
\ref{sec:second order} and \ref{sec:applications}.  Born-Infeld type
generalizations \cite{Gibbons1995} are also covered together
with first order manifestly duality invariant formulations
\cite{Deser:1976iy,Bunster:2010wv,Bunster:2011aw}, which fall into
this class when reformulated with suitable additional scalar fields
\cite{Barnich:2007uu}.

Consistent deformations of a gauge invariant action are deformations
that preserve the number (but not necessarily the form or the algebra)
of the gauge symmetries.  In the supergravity context, these are
called ``gaugings'', and the deformed theories are called ``gauged
supergravities'', even though the undeformed theories possess already
a gauge freedom. We shall often adopt this terminology here. We shall
consider only local deformations, i.e., deformations of the Lagrangian
by functions of the fields and their derivatives up to some finite
(but unspecified) order.

Gaugings in extended supergravities have a long history that goes back
to \cite{Freedman:1976aw,Fradkin:1976xz}.
 For maximal supergravity, the first gauging
has been performed in \cite{deWit:1981sst} in the Lagrangian
formulation of \cite{Cremmer:1979up}, which involves a specific choice
of so-called ``duality frame'' (a choice of ``electric" directions
among a set of electric-magnetic pairs). More recent gaugings
involving a change of the duality frame have been constructed in
\cite{Dall'Agata:2012bb}.  All these are reviewed in
\cite{Trigiante:2016mnt}.

These works consider from the very beginning deformations in which the
vector fields become Yang-Mills connections for a non-abelian
deformation of the original abelian gauge algebra.  The corresponding
couplings are induced through the replacement of the abelian
curvatures by non-abelian ones and the ordinary derivatives by
covariant ones, plus possible additional couplings necessary for
consistency.  One natural question to be asked is whether this
embraces all possible consistent deformations.  There exist of course
theorems establishing the uniqueness of the Yang-Mills coupling under
general conditions (see e.g. \cite{Deser:1963zzc,Barnich:1993pa}), but
couplings to nonlinear scalar fields were not considered in these
early works, which focused furthermore on algebra-deforming
deformations.

The gaugings of supergravities have revealed the importance of the
choice of duality frame, in the sense that the space of consistent
deformations depends on that choice (see
\cite{Andrianopoli:2002mf,Hull:2002cv} and the recent analysis in
\cite{Dall'Agata:2014ita,Inverso:2015viq}). In order to take this
feature into account, a formalism has been developed in
\cite{deWit:2002vt,deWit:2005ub,deWit:2007kvg,Samtleben:2008pe},
called the ``embedding tensor" formalism.  It is reviewed in \cite{Trigiante:2016mnt}.  
In this formalism,
additional fields are introduced besides those appearing in
(\ref{eq:1bis}), which are magnetic vector potentials and $2$-form
auxiliary gauge fields.  The theory possesses also additional gauge
symmetries.  The choice of duality frame is implicitly encoded in the
``embedding'' tensor, which is subject to a number of constraints.  It
was shown in \cite{Henneaux:2017kbx} that the space of consistent
deformations in the ``embedding" formalism is isomorphic to the space
of consistent deformations for the action (\ref{eq:1bis}) written in
the duality frame picked by the choice of embedding tensor.  For that
reason, one can investigate the question of gaugings by taking
(\ref{eq:1bis}) as starting point of the deformation procedure,
provided one allows the scalar field dependence in the vector piece of
the Lagrangian to cover all possible choices of duality frame.  It is
this task which is carried out here.  By doing so, one does not miss
any of the gaugings available in the embedding tensor formalism.

One systematic way to explore deformations of theories with a gauge
freedom is provided by the BV-BRST formalism \cite{Barnich:1993vg}.
In the BRST approach, inequivalent infinitesimal local gaugings
correspond to BRST cohomology classes in ghost number zero computed in
the space of local functionals. In this work, we completely
characterize the BRST cohomology for the theories defined by (\ref{eq:1bis}), i.e., we completely characterize, in four spacetime dimensions, the  
deformations of  abelian vector fields coupled non-minimally to scalar 
chargeless fields with a possibly non polynomial dependence on the (undifferentiated) scalar fields.

In particular, we show that besides the obvious deformations that
consist in adding gauge invariant terms to the Lagrangian without
changing the gauge symmetries, the gaugings can be related to the
global symmetries of the action (\ref{eq:1bis}).  These gaugings
modify the form of the gauge transformations.

The global symmetries can be classified into two different types: (i)
global symmetries with covariantizable Noether currents, where by
``covariantizable'', we mean that one can choose the ambiguities in
the Noether currents so as to take them gauge invariant ($V$-type
  symmetries); (ii) global symmetries with non-covariantizable
Noether currents. Only the first type directly gives rise to an
infinitesimal consistent deformation through minimal coupling of the
corresponding current to the vector potentials. 

The gaugings
associated with the other type of global symmetries need to satisfy
additional constraints.
This second type of global symmetries, in turn, can be subdivided into
two subtypes: (a) global symmetries with non-covariantizable Noether
currents that lead to a deformation that does not modify the gauge
algebra ($W$-type symmetries); (b) global symmetries with
non-covariantizable Noether currents that lead to a deformation that
does modify also the gauge algebra ($U$-type symmetries).  The
global symmetries of type (a) contain in their Noether current
non-gauge invariant Chern-Simons terms that cannot be removed by
suitably adjusting trivial contributions. The global symmetries of
type (b) are associated with
ordinary ``free" abelian gauge symmetries with co-dimension 2
conservation laws (see e.g.~\cite{Julia:1980gn} for an early
discussion).  The divergence of a current of type (a) is itself gauge invariant, while the divergence of a current of type (b) is not. Yang-Mills gaugings are associated with currents of type (b) and are hence of $U$-type. Topological couplings \cite{deWit:1987ph} are associated with non-covariantizable Noether currents of either type (a) or (b). ``Charging deformations" (if available), in which the scalar fields become charged but the gauge transformations of the vector fields  are not modified and remain therefore abelian, are of $V$- or $W$- type.

The BRST deformation procedure applies not only to the consistent
first order deformations, but also to higher orders where one might
encounter obstructions.  That procedure provides a natural
deformation-theoretic interpretation of quadratic constraints and
higher order constraints in terms of what is called the antibracket
map.

After establishing general theorems on the BRST cohomology valid without assuming a specific form of the Lagrangian or the rigid symmetries, including the above classification of the deformations and useful triangular properties of their algebra, we turn to various models that have been considered in the literature, for which we completely compute the deformations of $U$ and $W$-types. 

Our paper is organized as follows.  In Section \ref{sec:BRST}, we
provide a brief survey of the BRST deformation procedure. We then
compute in Section \ref{sec:gauging} the local BRST cohomology of the
models described by the action (\ref{eq:1bis}).  This is done by
following the method of \cite{Barnich:1994db,Barnich:1994mt} where the
BRST cohomology was computed for arbitrary compact -- in fact
reductive -- gauge group.  The difficulty in the computation comes
from the free abelian factors, where by ``free abelian factors'', we
mean abelian factors of the gauge algebra such that all matter fields
are uncharged, i.e., invariant under the associated gauge
transformations.  This is precisely the case relevant to the action
(\ref{eq:1bis}), which needs thus special care.  The method of
\cite{Barnich:1994db,Barnich:1994mt} is based on an expansion
according to the antifield number. It makes direct contact with
symmetries and conservation laws through the lowest antifield number
piece of the BRST differential, called the ``Koszul-Tate''
differential, which involves the equations of motion
\cite{Fisch:1990rp,Henneaux:1991rx}.  The Noether charges appear
through the ``characteristic'' cohomology, given by the local
cohomology of the ``Koszul-Tate'' differential \cite{Barnich:1994db}.

We then discuss in Section \ref{sec:AntiMap} the structure of the
antibracket map, which is relevant for the consistency of the
deformation at second order and the possible appearance of
obstructions, and provide information on the structure of the global
symmetry algebra.

The method of \cite{Barnich:1994db,Barnich:1994mt} provides the
general structure of the BRST cocycles in terms of conserved
currents. In order to reach more complete results, one must use
additional information specific to each model.  We therefore specify
further the models in Section \ref{sec:second order}, where we
concentrate on scalar-coupled second order Lagrangians that are
quadratic in the vector fields and their derivatives.  These
specialized models still cover the scalar-vector sectors of extended
supergravities.  Explicit examples are treated in detail to illustrate
the method in Section \ref{sec:applications}, where complete results
for the local BRST cohomology, up to the determination of $V$-type
symmetries, are worked out.  In Section \ref{sec:first-order-actions},
we then illustrate our techniques in the case of the manifestly
duality-symmetric first order action of \cite{Bunster:2011aw}, in the
formulation of \cite{Barnich:2007uu}, which is adapted to the direct
use of the methods developed here.

The last section (Section \ref{sec:conclusions}) summarizes our
results and recapitulates the structure of the local BRST cohomology.
Two appendices complete our work by respectively displaying our
notations and conventions on exterior forms and their duals (Appendix
\ref{sec:notations}) and discussing further properties of the
antibracket map (Appendix \ref{sec:antibr-maps-desc}).
Appendix C is devoted to the detailed analysis of the $W$-component of 
the commutator of two $U$-type transformations.

\section{BRST deformation theory: a quick survey}\label{sec:BRST}

\subsection{Batalin-Vilkovisky antifield formalism}

In order to systematically construct consistent interactions in gauge
theories, it is useful to reformulate the problem in the context of
algebraic deformation theory
\cite{Gerstenhaber:1964,nijenhuis1966cohomology,Julia:1986gv,Julia:1986ha}. The
appropriate framework is provided by the Batalin-Vilkovisky antifield
formalism
\cite{Batalin:1981jr,Batalin:1983jr,Barnich:1993vg,Henneaux:1997bm}.

The structure of an irreducible gauge system, i.e., the Lagrangian
$\cL_0$ with field content $\varphi^{a}$, generating set of gauge
symmetries\footnote{We use the condensed De Witt notation.}
$\delta_\epsilon \varphi^{a}=R^{a}{}_\alpha [\varphi^{b}]
\,(\epsilon^\alpha)$ and their algebra, is captured by the
Batalin-Vilkovisky (BV) master action $S$ (see
e.g.~\cite{Henneaux:1992ig,Gomis:1995he} for reviews). The master
action is a ghost number $0$ functional
\begin{equation}
  \label{eq:masteraction}
  S=\int \!d^n\!x\, \cL=\int \!d^n\!x\, \left[ \cL_0 +\varphi^*_{a}
    R^{a}{}_\alpha \,  
    (C^\alpha) + \frac{1}{2}C^*_\alpha
    f\indices{^\alpha_{\beta\gamma}}( C^\beta, C^\gamma) + \dots  \right], 
\end{equation}
that satisfies what is called the master equation
\begin{equation}
\label{eq:masterequation}
\half (S,S)=0.
\end{equation}
In this equation, the BV antibracket is the odd graded Lie bracket
defined by
\begin{equation}
  \label{eq:antibracket}
  (X,Y)= \int\!d^n\!x\, \left[ \frac{\delta^R X}{\delta \Phi^A(x)}
\frac{\delta^L Y}{\delta \Phi^*_A(x)}-\frac{\delta^R X}{\delta 
\Phi^*_A(x)}\frac{\delta^L Y}{\delta \Phi^A(x)} \right]
\end{equation}
on the extended space $\Phi^A=(\varphi^{a},C^\alpha,\dots)$ of
original fields and ghosts (and ghosts for ghosts in the case of
reducible gauge theories) and their antifields $\Phi^*_A$. The ghost
numbers of $\varphi^a,C^\alpha$ are $0,1$, while
${\rm gh}(\Phi^*_A)=-{\rm gh}(\Phi^A)-1$. The Lagrangian, gauge
variations and structure functions of the gauge algebra are contained
in the first, second and third term of the master action
\eqref{eq:masteraction} respectively.

For the deformation problem, one assumes the existence of an
undeformed theory described by $S^{(0)}$ satisfying the master
equation $\half (S^{(0)},S^{(0)})=0$ and one analyzes the conditions
coming from the requirement that, in a suitable expansion, the
deformed theory
\begin{equation}
  \label{eq:deformedmasteraction}
  S=S^{(0)}+ S^{(1)}+ S^{(2)}+\dots,
\end{equation}
satisfies the master equation \eqref{eq:masterequation}. The deformed
Lagrangian, gauge symmetries and structure functions can then be read
off from the deformed master action \eqref{eq:deformedmasteraction}.

The first condition on the ``infinitesimal'' deformation $S^{(1)}$ is
\begin{equation}
  \label{eq:40a}
  (S^{(0)},S^{(1)})=0.
\end{equation}
This equation admits solutions $S^{(1)}=(S^{(0)},\Xi)$, for all $\Xi$ of
ghost number $-1$.  Such deformations can be shown to be trivial in
the sense that they can be absorbed by (anticanonical) field-antifield
redefinitions. Moreover, trivial deformations in that sense are always
of the form $S^{(1)}=(S^{(0)},\Xi)$ for some local $\Xi$. It thus follows
that equivalence classes of deformations up to trivial ones are
classified by $H^0(s)$, the ghost number zero cohomology of the
antifield dependent BRST differential $s=(S^{(0)},\cdot)$ of the
undeformed theory,
\begin{equation}
  \label{eq:42}
  [S^{(1)}]\in H^0(s). 
\end{equation}
For our problem of determining the most general deformation, we start
by computing $H^{0}(s)$ and couple its elements with independent
parameters to the starting point action to obtain
$S^{(0)}+S^{(1)}$. The parameters thus play the role of generalized
coupling constants. In a second step, we determine the constraints on
these coupling constants coming from the existence of a completion
such that \eqref{eq:masterequation} holds. The expansion is then in
terms of homogeneity in these generalized coupling constants and not,
as often done, in homogeneity of fields (in which case $S^{(0)}$
corresponds to an action quadratic in the fields). In particular, this
approach treats the different types of symmetries involved in the
determination of $H^{0}(s)$ on the same footing.

In the standard field theoretic setting, one insists on spacetime
locality which implies that the cohomology is computed in the space of
local functionals in the fields and antifields. In turn, this can be
shown to be equivalent to the cohomology of $s$ in the space of local
functions up to total derivatives or, in form notation, to the
cohomology of $s$ in top form degree $n$, up to the horizontal
differential of an $n-1$ form. Local functions are functions that
depend on the spacetime coordinates, the fields and a finite number of
derivatives. The horizontal differential is not the de Rham
differential but is instead given by $d=dx^\mu\d_\mu$, where
$\d_\mu=\ddl{}{x^\mu}+\d_\mu z^\Sigma\ddl{}{z^\Sigma}+\dots$ is the
total derivative. Here, fields and antifields are collectively denoted
by $z^\Sigma=(\Phi^A,\Phi^*_A)$. More explicitly, the ghost number $g$
cohomology $H^g(s)$ of the antifield dependent BRST differential $s$
computed in the space of local functionals is isomorphic to
$H^{g,n}(s|d)$, where the latter group is defined by
\begin{equation}
  \label{eq:9a}
  s a^{g,n} +d a^{g+1,n-1}=0,\quad a^{g,n}\sim a^{g,n}+s b^{g-1,n}+d
  b^{g,n-1},
\end{equation}
the first superscript referring to ghost number and the second to form
degree.  The BRST differential is defined on the undifferentiated
fields and antifields by $s\Phi^A=-\vddr{\cL}{\Phi^*_A}$,
$s\Phi^*_A=\vddr{\cL}{\Phi^A}$. It is extended to the
derivatives through $[s,\d_\mu]=0$ resulting in $\{s,d\}=0$. This
reformulation allows one to use systematic homological techniques
(``descent equations'') for the computation of these classes (see
e.g.~\cite{DuboisViolette:1985jb}).

At second order, the condition on the infinitesimal deformation
$S^{(1)}$ is  
\begin{equation}
  \label{eq:43}
  \half (S^{(1)},S^{(1)})+(S^{(0)},S^{(2)})=0.
\end{equation}
The antibracket gives rise to a well defined map in cohomology,
\begin{equation}
  \label{eq:31a}
  (\cdot,\cdot): H^{g_1}(s|d)\otimes H^{g_2}(s|d)\longrightarrow H^{g_1+g_2+1}(s|d).
\end{equation}
For cocycles $C_i$ with $[C_i]\in H^{g_i}(s|d)$, it is explicitly given
by
\begin{equation}
  \label{eq:44}
  ([C_1],[C_2])=[(C_1,C_2)]\in H^{g_1+g_2+1}(s|d). 
\end{equation}
Condition \eqref{eq:43} constrains the infinitesimal
deformation $S^{(1)}$ to satisfy
\begin{equation}
  \label{eq:45}
  \half([S^{(1)}],[S^{(1)}])=[0]\in H^1(s|d).
\end{equation}
If this is the case, $S^{(2)}$ in \eqref{eq:43} is defined up to a
cocycle in ghost number $0$. Higher order brackets and constraints
can be analyzed in a similar way, see
e.g.~\cite{Retakh1993,Fuchs2001}.

Besides the group $H^0(s|d)$ that describes infinitesimal deformations,
and $H^1(s|d)$ that controls the obstructions to extending these to
finite deformations, one can furthermore show \cite{Barnich:1994db}
that $H^{g}(s|d)\simeq H^{n+g}_{\rm char}(d)$ for $g\leq -1\,$. The
latter ``characteristic'' cohomology groups are defined by forms
$\omega$ in the original fields $\varphi^a$ such that
\begin{equation}
  \label{eq:10}
  d\omega^{n+g}\approx 0,\quad 
  \omega^{n+g}\sim\omega^{n+g}+d \eta^{n+g-1}+t^{n+g},
\end{equation}
with $t^{n+g}\approx 0$ and where $\approx 0$ denote terms that vanish
on all solutions to the Euler-Lagrange equations of motion. In
particular, these groups can be shown to vanish for $g\leq -3$ in
irreducible gauge theories \cite{Barnich:1994db,Barnich:2000zw}. The
group $H^{-2}(s|d)$ describes equivalence classes of ``global"
reducibility parameters, i.e., particular local functions $f^\alpha$
such that $R\indices{^a_\alpha}(f^\alpha)\approx 0$ where
$f^\alpha\sim f^\alpha+t^\alpha$ with $t^\alpha\approx 0$.
This terminology reflects the fact that this cohomology may be non 
trivial even for (locally) ``irreducible" gauge systems, in other words in 
the absence of  $p$-form gauge fields with higher $p$.  This will become clear momentarily 
and is crucial in this paper.
  These classes
correspond to global symmetries of the master action rather than of
the original action alone \cite{Brandt:1996uv,Brandt:1997cz}. The
associated characteristic cohomology $H^{n-2}_{\rm char}(d)$ captures
non-trivial (flux) conservation laws. More generally in the case of
free abelian $p$-form gauge symmetry it was shown in
\cite{Julia:1980gn} that one can generalize the first Noether theorem
($p=0$) and deduce by a similar formula a class of
$H^{n-p-1}_{\rm char}(d)$ generalizing the electric flux which corresponds 
to the case $p=1$, i.e., to ordinary gauge invariance.
The groups $H^{-1-p}(s|d)$ appear for $p$-form gauge theories
and vanish for $p\ge 2$ in the irreducible case \cite{Henneaux:1996ws}.
The group
$H^{-1}(s|d)$ describes and generates the inequivalent global
symmetries, with $H^{n-1}_{\rm char}(d)$ encoding the associated
inequivalent Noether currents\footnote{This is the ``first" theorem by
  E. Noether and its converse. More details can be found in section
  6.1 of \cite{Barnich:2000zw}.}. 
We mention these groups here since
they play an important role in the determination of $H^0(s|d)$ as it
will be seen in Section \ref{sec:gauging} below.

When $g_1=-1=g_2$, $(\cdot,\cdot) : H^{-1}\otimes H^{-1}\to H^{-1}$; in
this case the antibracket map encodes the Lie algebra structure of the
inequivalent global symmetries \cite{Barnich:1996mr}. More generally,
it follows from $(\cdot,\cdot) : H^{-1}\otimes H^{g}\to H^{g}$ that,
for any ghost number $g$, the BRST cohomology classes form a
representation of the Lie algebra of inequivalent global symmetries.
As a side-remark, let us also mention that in the context of
perturbative quantum field theory, $H^1(s|d)$ classifies potential gauge
anomalies while $H^0(s|d)$ classifies counterterms.

For notational simplicity, we will drop the square brackets when
computing the antibracket map below, but keep in mind that it involves
classes and not their representatives.

\subsection{Depth of an element}
With any cocycle $\omega^{g,k}$ of the local BRST cohomology is associated a $(s,d)$-descent
\begin{equation}
  \label{eq:B1}
  s\omega^{g,k}_l+d\omega^{g+1,k-1}_l=0,\quad
  s\omega^{g+1,k-1}_l+d\omega^{g+2,k-2}_l=0,\, \dots \,,\,
  s\omega^{g+l,k-l}_l=0, 
\end{equation}
that stops at some BRST cocycle $\omega^{g+l,k-l}_l$. The length $l$ of the shortest non trivial
descent is called the ``depth'' of $[\omega^{g,k}]\in H^{g,k}(s|d)$.  The last element $\omega^{g+l,k-l}_l$ is then non trivial in 
$H^{g+l,k-l}(s)$.  The usefulness of the depth in analyzing the BRST cohomology is particularly transparent in  \cite{DuboisViolette:1985hc,DuboisViolette:1985jb,Dubois-Violette:1986cj}.

Local BRST cohomology classes $[\omega^{g,k}]\in H^{g,k}(s|d)$ are
thus characterized, besides ghost number $g$ and form degree $k$, by the
depth $l$. 
In appendix
\ref{sec:antibr-maps-desc}, we work out how the antibracket map
behaves with respect to the depth of its elements.

\section{Abelian vector-scalar models in 4
  dimensions}\label{sec:gauging}

\subsection{Structure of the models}
\label{sec:model}

We now apply the formalism to the scalar-vector models described by the action (\ref{eq:1bis}).
We write $ \mathcal{L}_0= \mathcal{L}_S[\phi^i]+\mathcal{L}_V[A^I_\mu,\phi^i]$.
  In four spacetime dimensions, there is no Chern-Simons term in the Lagrangian, which can be assumed to be strictly gauge invariant and not just invariant up to a total derivative.  
Gauge invariant functions are functions
that depend on $F^I_{\mu\nu}=\d_\mu A_\nu^I-\d_\nu A^I_\mu$, $\phi^i$
and their derivatives, but not on
$A_\mu^I,\d_{(\nu} A_{\mu)}^I,\d_{(\nu_1}\d_{\nu_2} A_{\mu)}^I$, etc.  Thus $\mathcal{L}_V[A^I_\mu,\phi^i]$ depends on the vector potentials $A^I_\mu$ only through $F^I_{\mu\nu}=\d_\mu A_\nu^I-\d_\nu A^I_\mu$
and their derivatives.

We define 
\begin{equation}
  \label{eq:47}
  \vddl{\mathcal L_V}{F^I_{\mu\nu}}=\half (\star G_I)^{\mu\nu}
\end{equation}
where the $(\star G_I)^{\mu\nu}$ are also manifestly gauge invariant functions.
The equations of motion
for the vector fields can be written as 
\begin{equation}
  \label{eq:4bis}
  \vddl{\mathcal L_0}{A_\mu^I}=\d_\nu(\star G_I)^{\mu\nu}
\end{equation}
and the Lagrangian can
be taken to be 
\begin{equation}
  \mathcal{L}_0= \mathcal{L}_S[\phi^i]+\mathcal{L}_V[A^I_\mu,\phi^i],
  \quad d^4x\, \mathcal{L}_V=\int_0^1\frac{dt}{t} [G_I F^I]
  [t A^I_\mu,\phi^i]. \label{4.4}
\end{equation}

The associated solution to the BV master 
equation is given by
\begin{equation}
  \label{eq:3bis}
  S^{(0)}=S_0+\int \!d^4\!x\, A^{*\mu}_I \d_\mu C^I.
\end{equation}
The ghost number of the various fields and antifields is
\begin{center}
\begin{tabular}{c|cccccc}
& $\phi^i$ & $A^I_\mu$ & $C^I$ & $\phi^*_i$ & $A^{*\mu}_I$ & $C^*_I$ \\ \hline
gh & $0$ & $0$ & $1$ & $-1$ & $-1$ & $-2$
\end{tabular}
\end{center}
and the action of the BRST differential is given by
\begin{align}
  & s \phi^i = 0, \quad s A^I_\mu = \d_\mu C^I, \quad s C^I = 0, \nn
  \\ & s \phi^*_i = \frac{\delta \cL_0}{\delta \phi^i}, \quad s
  A^{*\mu}_I = \d_\nu(\star G_I)^{\mu\nu}, \quad s C^*_I = - \d_\mu
  A^{*\mu} .
\end{align}

It is useful to introduce the antifield number
\begin{center}
\begin{tabular}{c|cccccc}
& $\phi^i$ & $A^I_\mu$ & $C^I$ & $\phi^*_i$ & $A^{*\mu}_I$ & $C^*_I$ \\ \hline
afd & $0$ & $0$ & $0$ & $1$ & $1$ & $2$
\end{tabular}
\end{center}
and the pure ghost number
\begin{center}
\begin{tabular}{c|cccccc}
& $\phi^i$ & $A^I_\mu$ & $C^I$ & $\phi^*_i$ & $A^{*\mu}_I$ & $C^*_I$ \\ \hline
pgh & $0$ & $0$ & $1$ & $0$ & $0$ & $0$
\end{tabular}
\end{center}
so that the ghost number is the difference between the pure ghost
number and the antifield number.

The BRST differential $s$ splits according to antifield number as \be
s = \delta + \gamma \ee where $\delta$ is the ``Koszul-Tate
differential'' \cite{Fisch:1990rp,Henneaux:1992ig} and has antifield
number $-1$.  The differential $\gamma$ has antifield number equal to
zero.  One has \be \delta^2 = 0, \; \; \; \delta \gamma + \gamma
\delta = 0, \; \; \; \; \gamma^2 = 0.  \ee The action of $\delta$ and
$\gamma$ are respectively given by
\begin{align}
  & \delta \phi^i = 0, \quad \delta A^I_\mu = 0, \quad \delta C^I = 0, \nn
  \\ & \delta \phi^*_i = \frac{\delta \cL_0}{\delta \phi^i}, \quad \delta
  A^{*\mu}_I = \d_\nu(\star G_I)^{\mu\nu}, \quad \delta C^*_I = - \d_\mu
  A^{*\mu} ,
\end{align}
and
\begin{align}
  & \gamma \phi^i = 0, \quad \gamma A^I_\mu = \d_\mu C^I, \quad \gamma
    C^I = 0 , \nn 
  \\ & \gamma \phi^*_i = 0, \quad \gamma
  A^{*\mu}_I = 0, \quad \gamma C^*_I = 0 .
\end{align}

In terms of the Koszul-Tate differential, the cocycle condition for
$m$ in
characteristic cohomology takes the form $ d m + \delta n = 0$.
This 
equation is the same as the (co)cycle condition for 
$n$ in the local
(co)homology of $\delta$, which is indeed $\delta n + dm = 0$.  Using
this observation, and vanishing theorems for $H(d)$ and $H(\delta)$ in
relevant degrees, one can establish isomorphisms between the
characteristic cohomology and $H(\delta \vert d)$
\cite{Barnich:1994db}. For example, the characteristic cohomology
  $H^{n-2}_{\rm char}(d)$ is given by the 2-forms $\mu^IG_I$, while
  $H^n_2(\delta \vert d)$ (where the superscript refers to form degree
  and the subscript to antifield number) is given by the 4-forms
  $d^4\! x \, \mu^I C^*_I$. The isomorphism is realized through the
  $(\delta,d)$-descent
\begin{equation}
\delta \, d^4x\, C^*_I+d \star\! A^*_I=0,\quad \delta \star \!A^*_I+d
G_I=0,
\end{equation}
where $A^*_I=dx^\mu A^*_{I\mu}$.

\subsection{Consistent deformations}

One can characterize the BRST cohomological classes with non trivial
antifield dependence in terms of conserved currents and rigid
symmetries for all values of the ghost number.  For definiteness, we
illustrate explicitly the procedure for $H^0(s \vert d)$ in maximum
form degree, which defines the local consistent deformations.  We
consider next the case of general ghost number.

The main equation to be solved for $a$ is 
\begin{equation}
\label{eq:cocycle}
sa + db = 0,
\end{equation}
 where $a$ has form degree
4 and ghost number $0$.  To solve it, we expand the cocycle $a$
according to the antifield number, \be a = a_0 + a_1 + a_2.  \ee 
Because $a$ has total ghost number zero, each term $a_n$ has antifield number $n$ and pure ghost number (degree in the ghosts) $n$ as well.
As shown in \cite{Barnich:1994mt}, the expansion stops at most at
antifield number $2$.  The term $a_0$ is the (first order) deformation
of the Lagrangian.  A non-vanishing $a_1$ corresponds to a deformation
of the gauge variations, while a non-vanishing $a_2$ corresponds to a
deformation of the gauge algebra. All three terms are related by the cocycle condition (\ref{eq:cocycle}).

\subsubsection{Solutions of $U$-type ($a_2$ non trivial)}

The first case to consider is when $a_2$ is non-trivial.  This defines
``class I'' solutions in the terminology of \cite{Barnich:1994mt},
which we call here ``$U$-type'' solutions to comply with the general
terminology introduced below.  One has from the general theorems of
\cite{Barnich:1994db,Barnich:1994mt} on the invariant characteristic
cohomology that \be a_2 = d^4 x\,C^*_I \Theta^I \ee with \be
\Theta^I=\frac{1}{2!}{f^I}_{J_1J_{2}} C^{J_1} C^{J_{2}} .  \ee 
Here
${f^I}_{J_1J_{2}}$ are some constants, antisymmetric in $J_1$, $J_2$.
The reason why the coefficient $d^4 x\,C^*_I $ of the ghosts in $a_2$
is determined by the characteristic cohomology follows from the
equation $\delta a_2 + \gamma a_1 + db_1 =0$ that $a_2$ must fulfill
in order for $a$ to be a cocycle of $H(s \vert d)$.  Given that $a_2$
has antifield number equal to $2$, it is the 
characteristic cohomology
in form degree $n-2=2$ that is relevant\footnote{The precise way to express the relation between the local cohomology of $\delta$
and the highest term of the equation obeyed by $a$ is given in
section 7 of \cite{Barnich:1994mt}: $a_2$ must be a non trivial representative 
of  $H_{inv}(\delta \vert d)$, more precisely it must come from 
$H_{inv,2}^4(\delta \vert d)$ in ghost number zero. This relates the U-type deformations 
to the free abelian factors of the undeformed gauge group.}.  We refer the reader to
\cite{Barnich:1994db,Barnich:1994mt} for the details.  The emergence
of the characteristic cohomology in the computation of $H(s \vert d)$
will be observed again for $a_1$ below, where it will be the conserved
currents that appear.  This central feature follows from the fact that
the Koszul-Tate differential, which encapsulates the equations of
motion, is an essential building block of the BRST differential. We must now find the 
lower terms $a_1+a_0$ and relate them as expected to Noether currents that 
correspond to $H_1^4(\delta \vert d)$.

By the argument of \cite{Barnich:1994mt} (section 8) suitably generalized (section 12), the term $a_1$ is then
found to be \be a_1 = \star A^*_IA^{K}\d_{K}\Theta^I + m_1 \ee
where $\gamma m_1 =0$ and $\d_K=\ddl{}{C^K}$.  
The term $m_1$ (to be determined by the next
equation) is linear in $C^I$ and can be taken to be linear in the
undifferentiated antifields $A^*_I$ and $\phi^*_i$ since derivatives
of these antifields, which can occur only linearly, can be redefined
away through trivial terms.  We thus write \be m_1 = \hat K=\star
A^*_I \hat g^I -\star \phi^*_i \hat \Phi^i \ee with \be \hat
g^I=dx^\mu g^{I}_{\mu K}C^{K}\;, \quad~ \hat \Phi^i=\Phi^{i}_{K}C^{K}.
\ee Here $g^{I}_{\mu K}$ and $\Phi^{i}_{K}$ are gauge invariant
functions, arbitrary at this stage, but which will be constrained by
the requirement that $a_0$ exists.

We must now consider the equation $\delta a_1 + \gamma a_0 + d b_0 = 0$
that determines $a_0$ up to a solution of $\gamma a'_0 + db'_0 =0$.
This equation is equivalent to
 \begin{equation}
 \label{eq:00}
  \left(\vddl{\mathcal L_0}{A_\mu^I}\delta_{K} A_\mu^I
  +\vddl{\mathcal
    L_0}{\phi^i}\delta_{K}\phi^i \right) C^K+\gamma \alpha_0 + \d_\mu \beta_0^\mu=0, 
\end{equation}
where we have passed to dual notations ($a_0 = d^4 x \,\alpha_0$,
$db_0 = d^4 x \,\d_\mu \beta_0^\mu$) and where we have set
\begin{equation}
 \delta_{K} A_\mu^I=A^J_\mu {f^I}_{JK}+g_{\mu
    K}^I  ,\quad \delta_{K}
  \phi^i=\Phi^i_{K}.  \label{eq:TypeISymmb}
\end{equation}
Writing $\beta_0^\mu = j^\mu_K C^K + $ ``terms containing derivatives
of the ghosts'', we read from (\ref{eq:00}), by comparing the
coefficients of the undifferentiated ghosts, that \be \vddl{\mathcal
  L_0}{A_\mu^I}\delta_{K} A_\mu^I +\vddl{\mathcal
  L_0}{\phi^i}\delta_{K}\phi^i + \d_\mu j_K^\mu=0 .  \label{eq:3.19} \ee A necessary
condition for $a_0$ (and thus $a$) to exist is therefore that
$ \delta_{K} A_\mu^I$ and $ \delta_K \phi^i$ define symmetries.
 
 To proceed further and determine $a_0$, we observe that the non-gauge
 invariant term $ \vddl{\mathcal L_0}{A_\mu^I} A^J_\mu {f^I}_{JK}$ in
 $\vddl{\mathcal L_0}{A_\mu^I}\delta_{K} A_\mu^I$ can be written as
 $ \partial_\mu \left( \star G^{\nu\mu}_I A_\nu^J {f^I}_{JK}\right) $
 plus a gauge invariant term, so that
 $j^\mu_K - \star G^{\mu\nu}_I A_\nu^J {f^I}_{JK}$ has a
 gauge invariant divergence. Results on the invariant cohomology of
 $d$ \cite{Brandt:1990gy,Dubois-Violette:1992ye} imply then that the
 non-gauge invariant part of such an object can only be a Chern-Simons
 form, i.e.
 $j^\mu_K - \star G^{\mu\nu}_I A_\nu^J {f^I}_{JK} = J^\mu_{K} + \half
 \epsilon^{\mu\nu\rho\sigma}A_\nu^I F_{\rho\sigma}^J h_{I|JK}$, or
\begin{equation}
  j^\mu_{K}
  =J^\mu_{K}+\star G^{\mu\nu}_I
  A_\nu^J {f^I}_{JK} + \half
  \epsilon^{\mu\nu\rho\sigma}A_\nu^I F_{\rho\sigma}^J 
  h_{I|JK}
\end{equation}  
where $J^\mu_{K}$ is gauge invariant and where the symmetries of the
constants $h_{I|JK}$ will be discussed in a moment.  It is useful to
point out that one can switch the indices $I$ and $J$ modulo a trivial
term.

The equation (\ref{eq:00}) becomes
$-(\partial_\mu j^\mu_K) \, C^K + \gamma \alpha_0 + \partial_\mu
\beta_0^\mu = 0$, i.e.,
$j^\mu_K \, (\gamma A_\mu^K) + \gamma \alpha_0 + \partial_\mu
{\beta'}_0^\mu = 0$.  The first two terms in the current yield
manifestly $\gamma$-exact terms,
\begin{equation}
J^\mu_K \,  (\gamma A_\mu^K) = \gamma(J^\mu_K \,   A_\mu^K), \; \;
\; \star G^{\mu\nu}_I 
A_\nu^J {f^I}_{JK} \, (\gamma A_\mu^K)= \frac12 \gamma (\star
G^{\mu\nu}_I A_\nu^J {f^I}_{JK} \, A_\mu^K)
\end{equation} 
and so $h_{I|JK}$ must be such that the term $A^I F^J dC^K h_{I|JK}$
is by itself $\gamma$-exact modulo $d$. This is a problem that has
been much studied in the literature through descent equations (see
e.g. \cite{Dubois-Violette:1986cj}).  It has been shown that
$h_{I|JK}$ must be antisymmetric in $J$, $K$ and should have vanishing
totally antisymmetric part in order to be ``liftable" to $a_0$ and
non-trivial,
\begin{equation}
  h_{I|JK}=h_{I|[JK]},\quad  h_{[I|JK]}=0. \label{eq:Symmh}
\end{equation}

Putting things together, one finds for $a_0$
\be
a_0 = A^I\d_I\hat J + \half G_I
  A^{K}A^{L}\d_{L}\d_{K}\Theta^I + \half F^I
  A^KA^L\d_L\d_K\Theta'_I
\ee
where
\be
\hat J=\star dx^\mu J_{\mu K}
    C^{K}\;,\quad \Theta'_I=\frac{1}{2}h_{I|J_1 J_2}C^{J_1} C^{ J_{2}} .
\ee

A non-trivial $U$-solution
modifies the gauge
algebra. Deformations of the Yang-Mills type belong to this class. A
$U$-solution is characterized by constants ${f^I}_{J_1J_{2}}$ which
are antisymmetric in $J_1$, $J_2$. These constants must be such that
there exist gauge invariant functions $g^{I}_{\mu K}$ and
$\Phi^{i}_{K}$ such that $ \delta_{K} A_\mu^I$ and $ \delta_K \phi^i$
define symmetries of the undeformed Lagrangian.  Here
$ \delta_{K} A_\mu^I$ and $ \delta_K \phi^i$ are given by
(\ref{eq:TypeISymmb}).  Furthermore, the $h$-term in the corresponding
conserved current (if any) must fulfill (\ref{eq:Symmh}).  The
deformation $a_0$ of the Lagrangian takes the Noether-like form.

Given the ``head'' $a_2$ of a $U$-type solution, characterized by a set of ${f^I}_{J_1J_{2}}$'s,  the lower terms $a_1$ and $a_0$, and in particular the $h$-piece, are not uniquely determined.  One can always add solutions of $W$, $V$ or $I$-types described below, which have the property that they have no $a_2$-piece.  
Hence one may require that the completion of the ``head'' $a_2$ of a $U$-type solution should be chosen to vanish when $a_2$ itself vanishes. But this leaves some freedom in the completion of $a_2$, since for instance any $W$-type solution multiplied by a component of ${f^I}_{J_1J_{2}}$ will vanish when the ${f^I}_{J_1J_{2}}$'s are set to zero. The situation has a triangular nature since two $U$-type solutions with the same $a_2$ differ by solutions of ``lower'' types, for which there might not be a canonical choice.
 
 Note that further constraints on ${f^I}_{J_1J_{2}}$ (notably the
 Jacobi identity) arise at second order in the deformation parameter.

 \subsubsection{Solutions of $W$ and $V$-type (vanishing $a_2$ but
   $a_1$ non trivial)}

These solutions are called ``class II'' solutions in \cite{Barnich:1994mt}.

We now have \be a = a_0 + a_1 \ee and $a_1$ can be taken to be gauge
invariant, i.e., annihilated by $\gamma$ \cite{Barnich:1994mt}.  We
thus have \be a_1 = \hat K=\star A^*_I \hat g^I -\star \phi^*_i \hat
\Phi^i \ee with \be \hat g^I=dx^\mu g^{I}_{\mu K}C^{K}\;, \quad~ \hat
\Phi^i=\Phi^{i}_{K}C^{K} \, . \ee Here $g^{I}_{\mu K}$ and $\Phi^{i}_{K}$
are again gauge invariant functions, which we still denote by the same
letters as above, although they are independent from the similar
functions related to the constants ${f^I}_{J_1J_{2}}$.  We also set
\begin{equation}
 \delta_{K} A_\mu^I=g_{\mu
    K}^I  ,\quad \delta_{K}
  \phi^i=\Phi^i_{K}.  \label{eq:DefKbb} 
\end{equation}

The equation $\delta a_1 + \gamma a_0 + d b_0 = 0$ implies then, as
above,
\begin{equation} \vddl{\mathcal L_0}{A_\mu^I}\delta_{K} A_\mu^I
  +\vddl{\mathcal L_0}{\phi^i}\delta_{K}\phi^i + \d_\mu j_K^\mu=0\,. \label{eq:3.28}
\end{equation}
A necessary condition for $a_0$ (and thus $a$) to exist is
therefore that $ \delta_{K} A_\mu^I$ and $ \delta_K \phi^i$ given by
(\ref{eq:DefKbb}) define symmetries. Equation (\ref{eq:3.28}) take the same form as Eq. (\ref{eq:3.19}), but there is one important difference: the divergence of the current  $j^\mu_{K}$ is now gauge invariant, while it is {\em not} in (\ref{eq:3.19}), due to the contribution coming from $a_2$.

The current takes this time the form
 \begin{equation}
  j^\mu_{K}
  =J^\mu_{K}+ \half
  \epsilon^{\mu\nu\rho\sigma}A_\nu^I F_{\rho\sigma}^J 
  h_{I|JK},
\end{equation}
(with $h_{I|JK}$ fulfilling the above symmetry properties) yielding
\be a_0 = A^I\d_I\hat J + \half F^I A^KA^L\d_L\d_K\Theta'_I \ee where still
\be \hat J=\star dx^\mu J_{\mu K} C^{K}\;,\quad
\Theta'_I=\frac{1}{2}h_{I|J_1 J_2}C^{J_1} C^{ J_{2}} .  \ee We define
$W$-type solutions to have $h_{I|JK} \neq 0$, while $V$-type have
$h_{I|JK}=0$.  Both these types deform the gauge transformations but
not their algebra (to first order in the deformation).  They are
determined by rigid symmetries of the undeformed Lagrangian 
with gauge invariant variations (\ref{eq:DefKbb}). The $V$-type have gauge invariant currents, while the
currents of the $W$-type contain a non-gauge invariant piece.

Note that again, the solutions of $W$ and $V$-types are determined up to a solution of lower type with no $a_1$-``head'', and that there might not be a canonical choice. In fact one may require similarly that $W$-type transformations become trivial when 
$h_{I|JK}$ tends to zero.

\subsubsection{Solutions of $I$-type (vanishing $a_2$ and $a_1$)}

In that case, 
\be
a = a_0
\ee
with $\gamma a_0 + db_0 =0$.

Since there is no Chern-Simons term in four dimensions, one can assume
that $b_0 =0$.  The deformation $b_0$ is therefore a gauge invariant
function, i.e., a function of the abelian curvatures $F_{\mu \nu}^I$,
the scalar fields, and their derivatives. The $I$-type deformations
neither deform the gauge transformations nor (a fortiori) the gauge
algebra.  Born-Infeld deformations belong to this type. They are
called ``class III'' solutions in \cite{Barnich:1994mt}.

\subsection{Local BRST cohomology at other ghost numbers}
\label{sec:locBRST}

\subsubsection{$h$-terms}

The previous discussion can be repeated straightforwardly at all ghost
numbers. The analysis proceeds as above.  The tools necessary to
handle the ``$h$-term'' in the non gauge invariant ``currents'' have
been generalized to higher ghost numbers through familiar means and
can be found in
\cite{DuboisViolette:1985hc,DuboisViolette:1985jb,Dubois-Violette:1986cj}.
 
The $h$-terms belong to the ``small'' or ``universal'' algebra
involving only the $1$-forms $A^I$, the $2$-forms $F^I = dA^I$, the
ghosts $C^I$ and their exterior derivative.  The product is the
exterior product.  One describes the $h$-term through a $(\gamma,d)$-descent
equation and what is called the ``bottom'' of that descent, which is
annihilated by $\gamma$ and has form degree $<4$ in four dimensions.
The only possibilities in the free abelian case are the $2$-forms \be
\frac1m h_{I \vert J_1 \cdots J_m} F^I C^{J_1} \cdots C^{J_m} \ee
where \be h_{I \vert J_1 \cdots J_m} = h_{I \vert [J_1 \cdots J_m]} .
\ee One can assume $h_{[I \vert J_1 \cdots J_m]} = 0$ since the
totally antisymmetric part gives a trivial bottom. The lift of this
bottom goes two steps, up to the $4$-form \be h_{I \vert J_1 J_2
  \cdots J_m} F^I F^{J_1}C^{J_2} \cdots C^{J_m} \ee producing along
the way  a $3$-form \be h_{I \vert J_1 J_2 \cdots J_m} F^I
A^{J_1}C^{J_2} \cdots C^{J_m}   \ee
which has the property of not being gauge (BRST) invariant although its exterior derivative is (modulo trivial terms). 

\subsubsection{Explicit description of cohomology}

By applying the above method, one finds that the local BRST cohomology
of the models of section \ref{sec:model} can be described along
exactly the same lines as given below.  Note that the cohomology at
negative ghost numbers reflect general properties of the
characteristic cohomology that go beyond the mere models considered
here \cite{Barnich:1994db}.

\begin{enumerate}[(i)]
\item $H^g(s|d)$ is empty for $g\leqslant -3$.  

\item $H^{-2}(s|d)$ is represented by the $4$-forms 
  \begin{equation}
    U^{-2}=\mu^I d^4x\, C^*_I\label{eq:11a}. 
\end{equation}
If $A^*_I=dx^\mu A^*_{I\mu}$, the associated descent equations
are
  \begin{equation}
s\, d^4x\, C^*_I+d \star A^*_I=0,\quad s \star A^*_I+d
G_I=0,\quad sG_I=0.\label{eq:14}
\end{equation}
Characteristic cohomology $H^{n-2}_{\rm char}(d)$ is then
represented by the 2-forms $\mu^IG_I$.

\item Several types of cohomology classes in ghost numbers $g \geqslant
  -1$, which we call $U$, $V$ and $W$-type, can be described by
  constants ${f^I}_{JK_1\dots K_{g+1}}$ which are antisymmetric in the
  last $g+2$ indices,
\begin{equation}
    \label{sk}
   {f^I}_{JK_1\dots K_{g+1}}={f^I}_{[JK_1\dots K_{g+1}]}, 
\end{equation}
and constants $h_{I|JK_1\dots K_{g+1}}$ that are antisymmetric in the last $g+2$
indices but without any totally antisymmetric part\footnote{We write
  $h_{IJ}:= h_{I|J}$ for $g=-1$.}, 
\begin{equation}
    \label{Y1}
  h_{I|JK_1\dots K_{g+1}}=h_{I|[JK_1\dots K_{g+1}]},\quad  h_{[I|JK_1\dots K_{g+1}]}=0, 
\end{equation}
together with gauge invariant
functions $g^I_{\mu K_1\dots K_{g+1}},\Phi^i_{K_1\dots K_{g+1}}$ 
that are antisymmetric in the last $g+1$ indices. They are constrained by
the requirement that the transformations
\begin{equation}
  \label{eq:16}
 \delta_{K_1\dots K_{g+1}} A_\mu^I=A^J_\mu {f^I}_{JK_1\dots K_{g+1}}+g_{\mu
    K_1\dots K_{g+1}}^I  ,\quad \delta_{K_1\dots K_{g+1}}
  \phi^i=\Phi^i_{K_1\dots K_{g+1}}, 
\end{equation}
define symmetries of the action in the sense that 
 \begin{equation}
  \label{eq:47A}
  \vddl{\mathcal L_0}{A_\mu^I}\delta_{K_1\dots K_{g+1}} A_\mu^I
  +\vddl{\mathcal
    L_0}{\phi^i}\delta_{K_1\dots K_{g+1}}\phi^i+\d_\mu j^\mu_{K_1\dots K_{g+1}}=0,
\end{equation}
with currents $j^\mu_{K_1\dots K_{g+1}}$ that are antisymmetric in the
last $g+1$ indices. This can be made more precise by making the gauge
(non-)invariance properties of these currents manifest. One finds
\begin{equation}\label{eq:current}
  j^\mu_{K_1\dots K_{g+1}}
  =J^\mu_{K_1\dots K_{g+1}}+\star G^{\mu\nu}_I
  A_\nu^J {f^I}_{JK_1\dots K_{g+1}} + \half
  \epsilon^{\mu\nu\rho\sigma}A_\nu^I F_{\rho\sigma}^J 
  h_{I|JK_1\dots K_{g+1}},
\end{equation}
where $J^\mu_{K_1\dots K_{g+1}}$ is gauge invariant and
antisymmetric in the lower $g+1$ indices. When taking into account that
\begin{equation}
  \label{eq:15}
  G_I F^J= d(G_I A^J+\star A^*_I C^J)+s(\star A^*_IA^J+d^4x C^*_IC^J) ,
  \quad F^IF^J=d(A^I F^J),  
\end{equation}
and defining $C^{{K_1}\dots K_g}=C^{K_1}\dots C^{K_g}$, 
\begin{equation}
  \begin{split}
    & \Theta^I=\frac{1}{(g+2)!}{f^I}_{J_1\dots J_{g+2}}C^{J_1\dots
      J_{g+2}},\qquad~ \Theta'_I=\frac{1}{(g+2)!}h_{I|J_1\dots
        J_{g+2}}C^{J_1\dots J_{g+2}}\;,\\
    & \hat J=\star dx^\mu J_{\mu K_1\dots K_{g+1}}
    \frac{1}{(g+1)!}C^{K_1\dots K_{g+1}}\;,\quad \hat K=(\star A^*_I
    \hat g^I -\star \phi^*_i \hat \Phi^i)\;,\\ 
    & \hat g^I=\frac{1}{(g+1)!}dx^\mu g^{I}_{\mu K_1\dots
      K_{g+1}}C^{K_1\dots K_{g+1}}\;, \quad~ \hat
    \Phi^i=\frac{1}{(g+1)!}\Phi^{i}_{K_1\dots
      K_{g+1}}C^{K_1\dots K_{g+1}}\;,
\label{eq:17b}
  \end{split}
\end{equation} 
the ``global symmetry" condition \eqref{eq:47A} is equivalent to a 
$(s,d)$-obstruction equation, 
\begin{equation}
  G_I
  F^J\d_J\Theta^I+F^IF^J\d_{J}\Theta'_{I}+s(\hat K +A^I\d_I \hat J) + d\hat J
 = 0,\label{eq:16b}
\end{equation}
with $\d_I=\ddl{}{C^I}$. Note that the last two terms combine into
\[d[\star dx^\mu J_{\mu K_1\dots K_{g+1}}]
  \frac{1}{(g+1)!}C^{K_1}\dots C^{K_{g+1}},\] so that this equation
involves gauge invariant quantities only. It is this form that arises
in a systematic analysis of the descent equations. One can now
distinguish the three types of solutions.

\begin{enumerate}[a)]
\item $U$-type corresponds to solutions with non vanishing
  ${f^I}_{JK_1\dots K_{g+1}}$ and particular
  ${}^U h_{I|JK_1\dots K_{g+1}}$, ${}^U g^I_{\mu K_1\dots K_{g+1}}$, ${}^U
  \Phi^i_{K_1\dots K_{g+1}}$, ${}^U J_{\mu K_1\dots K_{g+1}}$ that
  vanish when the $f$'s vanish (and that may be vanishing even when
  the $f$'s do not). As we explained above, different choices of the particular completion  ${}^U h_{I|JK_1\dots K_{g+1}}$, ${}^U g^I_{\mu K_1\dots K_{g+1}}$, ${}^U
  \Phi^i_{K_1\dots K_{g+1}}$, ${}^U J_{\mu K_1\dots K_{g+1}}$ of $a_2$ exist and there might not be a canonical one, but a completion exists if the $U$-type solution is indeed a solution. Similar ambiguity holds for the solutions of $W$ and $V$-types described below.  A $U$-type solution is trivial if and only
  if it vanishes. Denoting by $\hat K_U,\hat J_U,(\Theta'_U)_I$, the
  expressions as in \eqref{eq:17b} but involving the particular
  solutions, the associated BRST cohomology classes are represented by
\begin{multline}
  \label{eq:18b}
  U=(d^4x C^*_I +\star A^*_IA^{K}\d_{K}+ \half G_I
  A^{K}A^{L}\d_{L}\d_{K})\Theta^I \\+\hat K_U+\half F^I
  A^KA^L\d_L\d_K(\Theta'_U)_I+A^I\d_I\hat J_U,
\end{multline}
with $sU+d(\star
A^*_I\Theta^I+G_IA^J\d_J\Theta^I+F^IA^J\d_J(\Theta'_U)_I+\hat J_U)=0\,$;

\item $W$-type corresponds to solutions with vanishing $f$'s but non
vanishing $h_{I|J K_1\dots K_{g+1}}$ and particular ${}^W g^I_{\mu
  K_1\dots K_{g+1}},{}^W 
\Phi^i_{K_1\dots K_{g+1}}, {}^W J_{\mu K_1\dots K_{g+1}}$ that may be chosen to vanish
when the $h$'s vanish. Such solutions are trivial when the $h$'s
vanish. With the obvious notation, the associated BRST
cohomology classes are represented by 
\begin{equation}
  \label{eq:19b}
  W=\hat K_W+\half F^IA^KA^L\d_L\d_K\Theta'_I+A^I\d_I\hat J_W,
\end{equation}
with $sW+d(F^IA^J\d_J\Theta'_I+\hat J_W)=0\,$; 

\item $V$-type corresponds to solutions with vanishing $f$'s and
  $h$'s. They are represented by
\begin{equation}
  \label{eq:20}
  V=\hat K_V+A^I\d_I \hat J_V,
\end{equation}
with $s V+d\hat J_V=0\,$ and $s \hat J_V=0\,$.
$V$ and its descent have depth $1$.

\end{enumerate}

\item Lastly, $I$-type cohomology classes exist in ghost numbers
  $g\geqslant 0$ and are described by
\begin{equation}
\hat I=d^4x\, \frac{1}{g!} I_{K_1\dots K_g}C^{K_1}\dots C^{K_g}\label{eq:12}
\end{equation}
with $s\hat I=0$, i.e., gauge invariant $I_{K_1\dots K_g}$ that are
completely antisymmetric in the $K$ indices. Such classes are to be
considered trivial if the $I_{K_1\dots K_s}$ vanish on-shell up to a
total derivative. This can again be made more precise by making the
gauge (non-)invariance properties manifest: an element of type $I$ is
trivial if and only if
\begin{equation}
  \label{eq:13}
  d^4x\, I_{K_1\dots K_g}\approx dJ_{K_1\dots
    K_g}+{m^I}_{JK_1\dots K_g}G_IF^J+\half F^IF^J m'_{IJK_1\dots K_g},
\end{equation}
where $J_{K_1\dots K_g}$ are gauge invariant $3$ forms that are
completely antisymmetric in the $K$ indices, while
${m^I}_{JK_1\dots K_g},m'_{IJK_1\dots K_g}$ are constants that are
completely antisymmetric in the last $g+1$ indices. Note also that the
on-shell vanishing terms in \eqref{eq:13} need to be gauge invariant.
When there are suitable restrictions on the space of gauge invariant
functions (such as for instance $x^\mu$ independent, Lorentz invariant
polynomials with power counting restrictions) one may sometimes
construct an explicit basis of non-trivial gauge invariant $4$ forms,
in the sense that if $d^4x I\approx \rho^{\cA}I_{\cA}+d\omega^{3}$ and
$\rho^{\cA}I_{\cA}\approx d\omega^{3}$, then $\rho^{\cA}=0$. The
associated BRST cohomology classes are then parametrized by constants
${\rho^{\cA}}_{K_1\dots K_g}$.

\end{enumerate}

At a given ghost number $g\geqslant -1$, the cohomology is the direct sum
of elements of type $U,W,V$ and also $I$ when $g\geqslant 0$.

This completes our general discussion of the local BRST cohomology.
Reference \cite{Barnich:1994mt} also considered simple factors in
addition to the abelian factors, as well as any spacetime dimension
$\geq 3$. One can extend the above results to cover these cases. In a
separate publication \cite{Barnich2017}, the computation of the local
BRST cohomology $H^{*,*}(s \vert d)$ for gauge models involving general 
reductive gauge algebras will
be carried out by following the different route adopted in
\cite{Barnich:2000zw}, which did not consider free abelian factors in
full generality. As requested by the analysis of the deformations of
the action (\ref{eq:1bis}), reference \cite{Barnich2017} generalizes
Theorem 11.1 of \cite{Barnich:2000zw} to arbitrary reductive Lie
algebras that include also (free) abelian factors (and in any
spacetime dimension $\geq 3$).

\subsubsection{Depth of solutions}

The depth of the various BRST cocycles plays a key role in the analysis
of the higher-order consistency condition.  It is given here.

The $U$-type and $W$-type solutions have depth $2$ because they
involve $A_\mu j^\mu$ with a non-gauge invariant current.  The
$V$-type solutions have depth $1$ because the Noether term
$A_\mu j^\mu$ involves for them a gauge invariant current.  Finally,
$I$-type solutions clearly have depth $0$.

\section{Antibracket map and structure of symmetries}
\label{sec:AntiMap}

\subsection{Antibracket map in cohomology}
We now investigate the antibracket map
$H^g \otimes H^{g'} \to H^{g+g'+1}$ for the different types of
cohomology classes described above. It follows from the detailed
discussion of the cohomology in section \ref{sec:locBRST} that the
shortest non trivial length of descents, the ``depth'', of elements of
type $U,W,V,I$ is $2$, $2$, $1$, $0$. In particular, the antibracket map is
sensitive to the depth of its arguments: the depth of the map is
less than or equal to the depth of its most shallow element, see
Appendix \ref{sec:antibr-maps-desc}.

The antibracket map involving $U^{-2}=\mu^I d^4x\, C^*_I$ in $H^{-2}$ is given by
\begin{equation}
(\cdot,U^{-2}) : H^g \to H^{g-1}, \quad \omega^{g,n}\mapsto
\vddl{{}^R\omega^{g,n}}{C^I}\mu^I.
\label{eq:53A}
\end{equation}
More explicitly, it is trivial for $g=-2$. It is also trivial for
$g=-1$ except for $U$-type where it is described by
$f\indices{^I_J}\mapsto f\indices{^I_J}\mu^J$. For $g> 0$, it is
described by
$\rho^{\cA}_{K_1\dots K_g}\mapsto \rho^{\cA}_{K_1\dots K_g}\mu^{K_g}$
for $I$-type,
$k\indices{^{v_1}_{K_1\dots K_{g+1}}}\mapsto
k\indices{^{v_1}_{K_1\dots K_{g+1}}}\mu^{K_{g+1}}$ for $V$-type, 
$h_{IJK_1\dots K_{g+1}}\mapsto h_{IJK_1\dots K_{g+1}}\mu^{K_{g+1}}$
and
$f\indices{^I_{JK_1\dots K_{g+1}}} \mapsto f\indices{^I_{JK_1\dots
    K_{g+1}}}\mu^{K_{g+1}}$ for $U$- and $W$-type.

The antibracket map for $g, g' \geqslant -1$ has the following
triangular structure:
\begin{equation} \label{table}
\begin{array}{c|c|c|c|c}
  (\cdot,\cdot)  & U & W & V & I \\
  \hline
  U & U\oplus W\oplus V\oplus I & W\oplus V\oplus I  & V\oplus I & I
  \\
  \hline
  W & W\oplus V \oplus I & W\oplus V\oplus I & V\oplus I & I 
  \\
  \hline
  V & V\oplus I & V\oplus I & V\oplus I & I
\\
  \hline
  I & I & I & I & 0
\end{array}
\end{equation}
Indeed, $(\hat I,\hat I')=0$ because $I$-type cocycles can be chosen to be
antifield independent. For all other brackets involving $I$-type
cocycles, it follows from appendix \ref{sec:antibr-maps-desc} that the
result must have depth $0$ and the only such
classes are of $I$ type. Alternatively, since all cocycles can be
chosen to be at most linear in antifields, the result will be a
cocycle that is antifield independent and only classes of $I$-type
have trivial antifield dependence. It thus follows that $I$-type
cohomology forms an abelian ideal.

According to appendix \ref{sec:antibr-maps-desc}, the depth of the
antibracket map of $V$-type cohomology with $V,W,U$-type is
less or equal to $1$, so it must be of $V$- or $I$-type. 

Finally, the remaining structure follows from the fact that only
brackets of $U$-type cocycles with themselves may give rise to terms
that involve $C^*_I$'s.  

\subsection{Structure of the global symmetry
  algebra} \label{sec:globalsymmetries}

Let us now concentrate on brackets 
between two elements that
have both ghost number $-1$, i.e., on the detailed structure of the
Lie algebra of inequivalent global symmetries when taking into account
their different types.

In this case, one may use the table above supplemented by the fact
that $I^{-1}=0$.  Let then
\begin{equation}
U_{u}, \quad W_{w},\quad V_{v},
\end{equation} 
be bases of symmetries of $U,W,V$-type\footnote{These are bases in the
  cohomological sense, i.e., $\sum_u \lambda^u [U_u] = [0]$
  $\Rightarrow$ $\lambda^u = 0$ (and similarly for $W_w$ and
  $V_v$). In terms of the representatives, this becomes
  $\sum_{u} \lambda^u U_u = s a + db$ $\Rightarrow$ $\lambda^u =
  0$.}. At ghost number $g=-1$, equations \eqref{eq:18b},
\eqref{eq:19b}, \eqref{eq:20} give
\begin{equation}
  \label{eq:5bis}
  V_{v}=K_{v},\quad W_{w}=K_{w},\quad
  U_{u}= (f_{u})\indices{^I_J}[d^4x\, C^*_I C^J+\star
  A^*_I A^J]+K_{u}. 
\end{equation}

It follows from \eqref{table} that $V$-type symmetries and the direct
sum of $V$ and $W$-type symmetries form ideals in the Lie algebra of
inequivalent global symmetries.

The symmetry algebra $\mathfrak{g}_U$ is defined as the quotient of
all inequivalent global symmetries by the ideal of
$V\oplus W$-type symmetries. In particular, if $U$-type symmetries
form a sub-algebra, it is isomorphic to $\mathfrak g_U$.

First, $V$-type symmetries are parametrized by constants $k^v$,
$V^{-1}=k^v V_{v}$. The gauge invariant symmetry transformation on the
original fields then are
\begin{equation}
  \delta_{v}
  A^I_\mu=-(V_{v},A^I_\mu)=g\indices{_{{v}\mu}^I},\quad 
  \delta_{v} \phi^i=
  -(V_{v},\phi^i)=\Phi^i_{v}\label{eq:50}. 
\end{equation}
Furthermore, there
exist constants ${C^{v_3}}_{{v_1}{v_2}}$ such that
\begin{equation}
  \label{eq:25}
  ([V_{v_1}],[V_{v_2}]) = -{C^{v_3}}_{{v_1}{v_2}} [V_{v_3}]
\end{equation}
holds for the cohomology classes.
We choose the minus sign because 
\begin{equation}
  \label{eq:28a}
  (V_{v_1},V_{v_2}) = - d^4x (A^{*\mu}_I
  [\delta_{v_1},\delta_{v_2}] A_\mu^I+\phi^*_i
  [\delta_{v_1},\delta_{v_2}]\phi^i),
\end{equation}
so that the ${C^{v_3}}_{{v_1}{v_2}}$ are the structure constants of
the commutator algebra of the $V$-type symmetries,
$[\delta_{v_1},\delta_{v_2}] = {C^{v_3}}_{{v_1}{v_2}}
\delta_{v_3}$. For the functions $g\indices{_{v\mu}^I}$ and
$\Phi^i_v$, this gives
\begin{equation} \label{eq:commsymV}
  \begin{split}
    \delta_{v_1} g\indices{_{v_2\mu}^I} - \delta_{v_2}
    g\indices{_{v_1\mu}^I} &
    = {C^{v_3}}_{{v_1}{v_2}} g\indices{_{v_3\mu}^I} + (\text{trivial}) \\
    \delta_{v_1} \Phi^i_{v_2} - \delta_{v_2} \Phi^i_{v_1} &=
    {C^{v_3}}_{{v_1}{v_2}} \Phi^i_{v_3} + (\text{trivial}).
  \end{split}
\end{equation}
The ``trivial" terms on the right hand side take the form ``(gauge
transformation) $+$ (antisymmetric combination of the equations of
motion)" which is the usual ambiguity in the form of global
symmetries, see e.g. section 6 of \cite{Barnich:2000zw}. They come
from the fact that equation \eqref{eq:25} holds for classes: for the
representatives $V_v$ themselves, \eqref{eq:25} is
$(V_{v_1},V_{v_2}) = -{C^{v_3}}_{{v_1}{v_2}} V_{v_3} + sa + db$. The
trivial terms in \eqref{eq:commsymV} are then the symmetries generated
by the extra term $sa + db$, which is zero in cohomology.  The graded
Jacobi identity for the antibracket map implies the ordinary Jacobi
identity for these structure constants,
\begin{equation}
  \label{eq:26}
  {C^{v_1}}_{v_2[v_3}{C^{v_2}}_{v_4v_5]}=0. 
\end{equation}

Next, $W$-type symmetries are parametrized by constants $k^w$,
$W^{-1}=k^w W_{w}$ and encode the gauge invariant symmetry
transformations
\begin{equation}
  \delta_{w}
  A^I_\mu=-(W_{w},A^I_\mu)=g\indices{_{{w}\mu}^I},\quad 
  \delta_{w} \phi^i=
  -(W_{w},\phi^i)=\Phi^i_{w}\label{eq:49a} 
\end{equation}
with associated Noether 3 forms
$j_W=k^w (h_w)_{IJ}F^{(I}A^{J)} + k^w J_{Ww}$. There then exist
${C^{v_2}}_{w v_1}$, $C\indices{^{w_3}_{w_1 w_2}}$, $C\indices{^{v}_{w_1 w_2}}$
such that
\begin{equation}
  \label{eq:27a}
  \begin{split}
([W_w],[V_{v}]) &= -{C^{v_2}}_{w v} [V_{v_2}],\\
([W_{w_1}], [W_{w_2}]) &= - C\indices{^{w_3}_{w_1 w_2}} [W_{w_3}] -
C\indices{^{v}_{w_1 w_2}} [V_v],
\end{split}
\end{equation}
with associated Jacobi identities that we do not spell out. For the functions $g\indices{_{w\mu}^I}$ and $\Phi^i_w$, this implies
\begin{align}
    \delta_{w} g\indices{_{v\mu}^I} - \delta_{v} g\indices{_{w\mu}^I} &= {C^{v_2}}_{{w}{v}} g\indices{_{v_2\mu}^I}, \quad \delta_{w} \Phi^i_{v} - \delta_{v} \Phi^i_{w} = {C^{v_2}}_{{w}{v}} \Phi^i_{v_2}, \\
    \delta_{w_1} g\indices{_{w_2\mu}^I} - \delta_{w_2}
    g\indices{_{w_1\mu}^I} &= {C^{w_3}}_{{w_1}{w_2}}
    g\indices{_{w_3\mu}^I} + {C^{v}}_{{w_1}{w_2}}
    g\indices{_{v\mu}^I}, \label{eq:435}
    \\
    \delta_{w_1} \Phi^i_{w_2} - \delta_{w_2} \Phi^i_{w_1} &=
    {C^{w_3}}_{{w_1}{w_2}} \Phi^i_{w_3} + {C^{v}}_{{w_1}{w_2}}
    \Phi^i_{v}\,,
    \label{eq:436}
  \end{align}
up to trivial terms, see the discussion below \eqref{eq:commsymV}.

Finally, $U$-type symmetries are parametrized by $k^{u}$,
$U^{-1}=k^{u} U_{u}$ and encode the symmetry
transformations
\begin{equation}
  \begin{split}
  \label{eq:51a}
  \delta_{u} A^I_\mu &= -(U_{u},A^I_\mu)=(f_{u})\indices{^I_J}A^J_\mu
  + g\indices{_{{u}\mu}^I},\quad \delta_{u} \phi^i=
  -(U_{u},\phi^i)=\Phi^i_{u}, \\
  \delta_{u} A^{*\mu}_I &= -(f_{u})\indices{^K_I} A^{*\mu}_K
  - \frac{\delta}{\delta A^I_\mu} ( A^{*\nu}_K g\indices{_{u\nu}^K} +
  \phi^*_i \Phi^i_u ),
  \quad \delta_{u} \phi^*_i = - \frac{\delta}{\delta \phi^i}
  ( A^{*\nu}_K g\indices{_{u\nu}^K} + \phi^*_j \Phi^j_u ),  \\
  \delta_{u} C^I &= (f_{u})\indices{^I_J} C^J,\quad \delta_{u} C^*_I =
  - (f_{u})\indices{^K_I} C^*_K.
\end{split}
\end{equation}
Again, there exist constants $C$ with various types of indices such that
\begin{align}
    ([U_u],[V_{v}]) &= -C\indices{^{v_2}_{u v}} [V_{v_2}],
    \label{eq:UVbracket}
    \\
([U_u],[W_{w}]) &= -C\indices{^{w_2}_{u w}} [W_{w_2}] -C\indices{^{v}_{u w}} [V_{v}],
\label{eq:UWbracket}
\\
([U_{u_1}], [U_{u_2}]) &= -C\indices{^{v}_{u_1 u_2}} [V_{v}]
-C\indices{^{w}_{u_1 u_2}} [W_{w}] -C\indices{^{u_3}_{u_1 u_2}} [U_{u_3}],
                     \label{eq:29b}
\end{align}
with associated Jacobi identities. Working
out the term proportional to $C^*_I$ in
$(U_{u_1},U_{u_2})$ gives the commutation relations for
the $(f_u)\indices{^I_J}$ matrices,
\begin{equation}\label{eq:commU}
[f_{u_1},f_{u_2}] = -C\indices{^{u_3}_{u_1u_2}}
f_{u_3} .
\end{equation}
In turn, this implies Jacobi identities for this type of structure
constants alone:
\begin{equation}
  \label{eq:26a}
  {C^{u_1}}_{u_2[u_3}{C^{u_2}}_{u_4u_5]}=0. 
\end{equation}
The $C\indices{^{u_3}_{u_1u_2}}$ are the structure constants of
  $\mathfrak{g}_U$. 

From equation \eqref{eq:UVbracket}, we get the identities
\begin{align}\label{eq:422}
  \delta_{u} g\indices{_{v\mu}^I} - \delta_{v} g\indices{_{u\mu}^I}
  - (f_u)\indices{^I_J} g\indices{_{v\mu}^J} &= {C^{v_2}}_{{u}{v}}
  g\indices{_{v_2\mu}^I}, \quad \delta_{u} \Phi^i_{v}
 - \delta_{v} \Phi^i_{u} = {C^{v_2}}_{{u}{v}} \Phi^i_{v_2} .
\end{align}
Equation \eqref{eq:UWbracket} gives the same identities with the
right-hand side replaced by the appropriate sum, as in
\eqref{eq:435}--\eqref{eq:436}. The last relation \eqref{eq:29b} gives
\begin{equation}\label{eq:423}
  \begin{split}
  \delta_{u_1} g\indices{_{u_2\mu}^I} - (f_{u_1})\indices{^I_J}
  g\indices{_{u_2\mu}^J}
  - (u_1 \leftrightarrow u_2) &= {C^{u_3}}_{{u_1}{u_2}}
  g\indices{_{u_3\mu}^I}
  + {C^{w}}_{{u_1}{u_2}} g\indices{_{w\mu}^I} + {C^{v}}_{{u_1}{u_2}}
  g\indices{_{v\mu}^I}, \\
  \delta_{u_1} \Phi^i_{u_2} - \delta_{u_2} \Phi^i_{u_1} &=
  {C^{u_3}}_{{u_1}{u_2}} \Phi^i_{u_3}
  + {C^{w}}_{{u_1}{u_2}} \Phi^i_{w} + {C^{v}}_{{u_1}{u_2}} \Phi^i_{v} .
\end{split}
\end{equation}
Equations \eqref{eq:422} and \eqref{eq:423} are again valid only
  up to trivial symmetries.

Let us now concentrate on identities containing the $h_{IJ}$, which
appear in the currents of $U$ and $W$-type. We first consider
$(U_u,W_w)$ projected to $W$-type. As in appendix
\ref{sec:antibr-maps-desc}, we have
$s(U_u,W_w)_{{\rm alt}}=-d(U_u,(h_{w})_{IJ}F^{I}A^{J}+J_{w})_{{\rm
    alt}}=d\{(h_{w})_{IJ} [(f_u)\indices{^I_K} F^K A^J +F^I
(f_u)\indices{^J_K} A^K ]+{\rm invariant}\}$. When comparing this to
$s$ applied to the right hand side of \eqref{eq:UWbracket} and using
the fact that $W$-type cohomology is characterized by the Chern-Simons
term in its Noether current, we get
\begin{equation} \label{eq:hwfu}
(h_{w})_{IN}(f_u)\indices{^I_M}+(h_{w})_{MI}(f_u)\indices{^I_N}=
C\indices{^{w_2}_{u w}}(h_{w_2})_{MN}. 
\end{equation}
This computation amounts to identifying the Chern-Simons term in
  the $U$-variation $\delta_u j_w$ of a current of $W$-type.  The same
  computation applied to $(W_{w_1}, W_{w_2})$ shows that
  $C\indices{^{w_3}_{w_1 w_2}}(h_{w_3})_{MN} = 0$, which implies
\begin{equation} \label{eq:hwfw}
  C\indices{^{w_3}_{w_1 w_2}} = 0
\end{equation}
since the matrices $h_w$ are linearly independent (otherwise, the
$W_w$ would not form a basis). In other words, the $W$-variation
$\delta_{w_1} j_{w_2}$ of a current of $W$-type is gauge invariant up
to trivial terms, i.e., is of $V$-type. 

In order to work out
$(U_{u_1},U_{u_2})$ projected to $W$-type, a slightly involved
reasoning gives 
\begin{equation}
\label{eq:Utrasf}
\delta_u G_I + (f_u)\indices{^J_I} G_J \approx  - 2 (h_u)_{IJ} F^J + \lambda^w_u (h_w)_{IJ} F^J + d(\text{invariant})
\end{equation}
for some constants $\lambda^w_u$. This is proved in Appendix \ref{app:derivation}  in the case where $G_I$ does not depend on derivatives of $F^I$ (but can have otherwise arbitrary dependence of $F^I$). We were not able to find the analog of \eqref{eq:Utrasf} in the higher derivative case.

Applying then $(U_{u_1},\cdot)_{\rm alt}$
to the chain of descent equations for $U_{u_2}$ and adding the chain
of descent equations for $C^{u_3}_{u_1u_2} U_{u_3}$ yields
\begin{align}
  (h_{u_2})_{IN}(f_{u_1})\indices{^I_M} &+(h_{u_2})_{MI}(f_{u_1})\indices{^I_N}
  -
    (h_{u_1})_{IN}(f_{u_2})\indices{^I_M}-(h_{u_1})_{MI}(f_{u_2})\indices{^I_N}
    \nonumber \\ &+ \frac{1}{2} \left[ (h_w)_{IN} (f_{u_2})\indices{^I_M} +  (h_w)_{IM} (f_{u_2})\indices{^I_N}  \right]\lambda^w_{u_1} \nonumber \\
&= C\indices{^{u_3}_{u_1 u_2}}(h_{u_3})_{MN} + C\indices{^{w}_{u_1
                   u_2}}(h_{w})_{MN}.
\end{align}
Again, this amounts to identifying the Chern-Simons terms in the
$U$-variation $\delta_{u_1} j_{u_2}$ of a $U$-type current. Equation
\eqref{eq:Utrasf} is crucial for this computation since $U$-type
currents contain $G_I$.
Using \eqref{eq:hwfu}, this becomes
\begin{align}\label{eq:hufu}
  (h_{u_2})_{IN}(f_{u_1})\indices{^I_M} &+(h_{u_2})_{MI}(f_{u_1})\indices{^I_N}
  -
    (h_{u_1})_{IN}(f_{u_2})\indices{^I_M}-(h_{u_1})_{MI}(f_{u_2})\indices{^I_N}
    \nonumber \\ &= C\indices{^{u_3}_{u_1 u_2}}(h_{u_3})_{MN} + \left[ C\indices{^{w}_{u_1 u_2}} - \frac{1}{2} C\indices{^w_{u_2 w_2}} \lambda^{w_2}_{u_1} \right] (h_{w})_{MN}.
\end{align}
We see that the effect of the $\lambda^w_u$ is to shift the structure constants of type $C\indices{^{w}_{u_1 u_2}}$. The constants $\lambda^w_u$ vanish for the explicit models considered below; it would be interesting to find an explicit example where this is not the case. As a last comment, we note that antisymmetry of equation \eqref{eq:hufu} in $u_1$ and $u_2$ imposes the constraint
\begin{equation}
C\indices{^w_{u_2 w_2}} \lambda^{w_2}_{u_1} + C\indices{^w_{u_1 w_2}} \lambda^{w_2}_{u_2} = 0
\end{equation}
on the constants $\lambda^w_u$.

\subsection{Parametrization through symmetries} \label{sec:parametrization}

It follows from the discussion of the antibracket map involving
$H^{-2}$ after \eqref{eq:53A} that cohomologies of $U,W,V$-type in
ghost numbers $g\geqslant 0$ can be parametrized by symmetries of the
corresponding type with suitably constrained coefficients
\begin{equation}
  \label{gd}k\indices{^{u}_{K_1\dots K_{g+1}}},
  \quad k\indices{^{v}_{K_1\dots K_{g+1}}},\quad k\indices{^{w}_{K_1\dots K_{g+1}}}.
\end{equation}
In this way, for $g=0$, the problem of finding all infinitesimal
gaugings can be reformulated as the question of which of these
symmetries can be gauged.

In order to do this, it is useful to first rewrite the
$h_{I|JK_1\dots K_{g+1}}$ appearing in the cohomology classes of $U$
and $W$-types in the equivalent symmetric convention
\begin{equation}
    X_{IJ,K_1\dots K_{g+1}}:=h_{(I|J)K_1\dots K_{g+1}}\iff 
    h_{I|JK_1\dots K_{g+1}}=\frac{2(g+2)}{g+3}X_{I[J,K_1\dots K_{g+1}]}
\label{Xintermsofh}
\end{equation}
where \eqref{Y1} is now replaced by
\begin{equation}\label{YX}
  X_{IJ,K_1 \dots K_{g+1}} = X_{(IJ),[K_1 \dots K_{g+1}]}, \quad
  X_{(IJ,K_1) K_2 \dots K_{g+1}} = 0\;. \end{equation}
Note that for $g=-1$, $h_{IJ} = X_{IJ}$.

For cohomology classes of $U,W$-type, we can write
\begin{align}
  \label{eq:352}
  {f^I}_{JK_1\dots K_{g+1}} &= (f_{u})\indices{^I_{J}}
                              \, k\indices{^{u}_{K_1\dots K_{g+1}}}, \\
  {}^U X_{IJ,K_1\dots K_{g+1}} &=(h_{u})_{IJ} \, k\indices{^{u}_{K_1\dots K_{g+1}}}, \\
  X_{IJ,K_1\dots K_{g+1}} &= (h_{w})_{IJ} \, k\indices{^{w}_{K_1\dots K_{g+1}}},
\end{align}
where $(f_{u})\indices{^I_{J}}$, $(h_{u})_{IJ}$ and $(h_{w})_{IJ}$
appear in the basis elements $U_{u}$ and $W_{w}$. (One has similar
parametrizations for the quantities $g^I_{\mu K_1\dots K_{g+1}}$,
$\Phi^i_{K_1\dots K_{g+1}}$, $J_{\mu K_1\dots K_{g+1}}$ in the
cohomology classes of the various types.)  This guarantees that
condition \eqref{eq:47A} (or \eqref{eq:16b}) is automatically
satisfied.

However, the symmetry properties \eqref{sk} and \eqref{YX} imply the
following linear constraints on the parameters:
\begin{align}
(f_{u})\indices{^I_{(J}} \, k\indices{^{u}_{K_1)K_2\dots K_{g+1}}} &= 0, \label{linf}\\
(h_{u})_{(IJ} \, k\indices{^{u}_{K_1)K_2 \dots K_{g+1}}} &= 0, \label{358}\\
(h_{w})_{(IJ} \, k\indices{^{w}_{K_1) K_2 \dots K_{g+1}}} &= 0. \label{465}
\end{align}
From the discussion of the cohomology, it also follows that $V$-type
cohomology classes are entirely determined by $V$-type symmetries in
terms of $k\indices{^{v}_{K_1\dots K_{g+1}}}$ without any additional
constraints.

\subsection{2nd order constraints on deformations and gauge algebra}
\label{sec:parametrization-2nd-order}

The most general infinitesimal gauging is given by
$S^{(1)}= \int ( U^0+W^0+V^0+I^0 )$. We have
\begin{equation}
\half (S^{(1)}, S^{(1)}) = \int \left( U^1+W^1+V^1+I^1 \right) \label{eq:22a}. 
\end{equation}
The infinitesimal deformation $S^{(1)}$ can be extended to second
order whenever the right hand side vanishes in cohomology, resulting in quadratic
constraints on the constants $k\indices{^{u_1}_K}$,
$k\indices{^{w_1}_K}$, $k\indices{^{v_1}_K}$ and $\rho^{\cA}$.
Working all of them out explicitly requires computing all brackets
between $U^0$, $W^0$, $V^0$ and $I^0$.

However, it follows from the previous section that the only
contribution to $U^1$ comes from $\half (U^0,U^0)$. The vanishing of
the terms containing the antighosts $C^*_I$ requires
\begin{equation}
  \label{eq:23}
  f\indices{^I_{J[K_1}}f\indices{^J_{K_2K_3]}}=0,
\end{equation}
i.e., the Jacobi identity for the $f\indices{^I_{JK}}$. The
  associated $n_v$-dimensional Lie algebra is the gauge
  algebra and is denoted by $\mathfrak g_g$.

Using
$f\indices{^I_{JK}} = (f_{u_1})\indices{^I_J} k\indices{^{u_1}_K}$ and
equation \eqref{eq:commU}, the Jacobi identity reduces to the
following quadratic constraint on $k\indices{^{u_1}_K}$:
\begin{equation} \label{eq:quadu}
 k\indices{^{u_1}_I} k\indices{^{u_2}_J} C\indices{^{u_3}_{u_1u_2}} -
  (f_{u_4})\indices{^K_I} k\indices{^{u_4}_J} k\indices{^{u_3}_K} = 0 .
\end{equation}
Note that the antisymmetry in $IJ$ of the second term is guaranteed by
the linear constraint \eqref{linf}. 
The terms at antifield number $1$ give the constraints 
\begin{align}
 \delta_I g^K_J + f\indices{^K_{MJ}} g^M_I - (I\leftrightarrow J) &= f\indices{^L_{IJ}} g^K_L \\
 \delta_I \Phi^i_J  - (I\leftrightarrow J) &= f\indices{^L_{IJ}} \Phi^i_L.
\end{align}
Expressed with $k$'s, this gives
\begin{equation}
    k\indices{^{\Gamma}_I} k\indices{^{\Delta}_J} C\indices{^{\Sigma}_{\Gamma\Delta}} -
  (f_{u})\indices{^K_I} k\indices{^{u}_J} k\indices{^{\Sigma}_K} = 0,
\end{equation}
where the capital Greek indices take all values $u,w,v$.  This gives
three constraints, according to the type of the free index
$\Sigma$. When $\Sigma = u$, we get the constraint $\eqref{eq:quadu}$,
because the only non-vanishing structure constants with an upper $u$
index are the $C\indices{^{u_3}_{u_1u_2}}$. When $\Sigma = w$, the
possible structure constants are $C\indices{^{w_3}_{w_1w_2}}$,
$C\indices{^{w_3}_{u_1w_2}} = -C\indices{^{w_3}_{w_2u_1}}$ and
$C\indices{^{w_3}_{u_1u_2}}$, giving the constraint
\begin{equation}
  k\indices{^{w_1}_I} k\indices{^{w_2}_J} C\indices{^{w_3}_{w_1w_2}}
  + 2 k\indices{^{u_1}_{[I}} k\indices{^{w_2}_{J]}}
  C\indices{^{w_3}_{u_1w_2}}
  + k\indices{^{u_1}_I} k\indices{^{u_2}_J} C\indices{^{w_3}_{u_1u_2}}
  - (f_{u_4})\indices{^K_I} k\indices{^{u_4}_J} k\indices{^{w_3}_K} = 0 .
\end{equation}
When the free index $\Sigma$ is of type $v$, one gets a similar
identity with all possible types of values in the lower indices of the
structure constants,
\begin{multline}
  k\indices{^{v_1}_I} k\indices{^{v_2}_J} C\indices{^{v_3}_{v_1v_2}} +
  2 k\indices{^{w_1}_{[I}} k\indices{^{v_2}_{J]}}
  C\indices{^{v_3}_{w_1v_2}} + k\indices{^{w_1}_I} k\indices{^{w_2}_J}
  C\indices{^{v_3}_{w_1w_2}} + 2 k\indices{^{u_1}_{[I}}
  k\indices{^{v_2}_{J]}} C\indices{^{v_3}_{u_1v_2}} \\ + 2
  k\indices{^{u_1}_{[I}} k\indices{^{w_2}_{J]}}
  C\indices{^{v_3}_{u_1w_2}} + k\indices{^{u_1}_I} k\indices{^{u_2}_J}
  C\indices{^{v_3}_{u_1u_2}} - (f_{u_4})\indices{^K_I}
  k\indices{^{u_4}_J} k\indices{^{v_3}_K} = 0 . 
\end{multline}

\section{Quadratic vector models}
\label{sec:second order}

\subsection{Description of the model}
\label{sec:lagrangian}

To go further, one needs to specialize the form of the Lagrangian,
which has been assumed to be quite general so far.  In this section,
we focus on second order Lagrangians arising in the context of
supergravities that contain $n_s$ scalar fields and depend
quadratically on $n_v$ abelian vector fields, non-minimally coupled to
each other, in four space-time dimensions.

More specifically, we take $\cL = \cL_S + \cL_V$, where
\begin{equation} 
  \cL_V = - \frac{1}{4}\, \mathcal{I}_{IJ}(\phi) F^I_{\mu\nu}
          F^{J\mu\nu}
          + \frac{1}{8} \,\mathcal{R}_{IJ}(\phi)\,
          \varepsilon^{\mu\nu\rho\sigma}
          F^I_{\mu\nu} F^J_{\rho\sigma} \label{eq:lag}
\end{equation}
and the scalar Lagrangian is of the sigma model form
\begin{equation}
  \label{eq:22}
  \mathcal L_S=-\half g_{ij}(\phi)\d_\mu\phi^ i\d^\mu\phi^j-V(\phi)
\end{equation}
where $g_{ij}$ is symmetric and invertible. Both $g_{ij}$ and $V$ depend only on undifferentiated scalar fields. Neglecting gravity,
this is the generic bosonic sector of ungauged supergravity. The
symmetric matrices $\cI$ and $\cR$, with $\cI$ invertible, depend only on
undifferentiated scalar fields and encode the non-minimal
couplings between the scalars and the abelian vectors.
The Bianchi identities and equations of motion for the vector fields
are given by
\begin{equation} \label{eq:eom}
  \partial_\mu (\star F^I)^{\mu\nu} = 0, \qquad \partial_\nu (\star G_I)^{\mu\nu} \approx 0.
\end{equation}
The Lagrangian
\eqref{eq:lag} falls into the general class of models described
previously, with the gauge invariant two-form
$G_I=\mathcal I_{IJ}\star F^J+\mathcal R_{IJ} F^J$ and
$d^4x\, \mathcal{L}_V=\frac{1}{2} G_I F^I$.

We assume
\begin{equation}
  \label{eq:33}
  \mathcal R_{IJ}(0)=0. 
\end{equation}
Note that a constant part in $\mathcal R_{IJ}$ can be put to zero
without loss of generality since the associated term in the Lagrangian
is a total derivative. In most cases, we also take 
$V=0$ or assume (writing $\d_i=\ddl{}{\phi^i}$) that
\begin{equation}
  \label{eq:34}
  (\d_iV)(0)=0. 
\end{equation}

\subsection{Constraints on $U$, $W$-type symmetries}
\label{sec:u-w-type}

We assume here and in the examples below that there is no explicit
$x^\mu$-dependence in the space of local functions in order to
constrain $U$ and $W$-type symmetries. For simplicity, we also assume that the potential vanishes, $V=0$. 
In section \ref{sec:applications}, these constraints will allow us to determine all symmetries of $U$ and $W$ type for specific models.

We need the scalar field equations, which are encoded in
\begin{equation}
s\star \phi^*_i+d(g_{ij} \star d\phi^j)=-\star
  \d_i(\mathcal{L}_S+\mathcal{L}_V)\label{eq:24},
\end{equation}
where $\d_i=\ddl{}{\phi^i}$. For $g=-1$, equation \eqref{eq:16b}
becomes
\begin{equation}
  \label{eq:17a}
  G_I F^J{f^I}_J+F^IF^Jh_{IJ}+dI^{n-1}-dG_I\, g^{I}-[d(g_{ij} \star
  d\phi^j)+\star
  \d_i(\mathcal{L}_S+\mathcal{L}_V)]\Phi^{i}
=0.
\end{equation}
When putting all derivatives of $F^I_{\mu\nu},\phi^i$ to zero, one
remains with
\begin{equation}
  \label{eq:18a}
  G_I F^J{f^I}_J+F^IF^Jh_{IJ}
-\star \d_i \mathcal{L}_V\, 
\Phi^{i}|_{{\rm der}=0}=0.
\end{equation}
It is here that the assumption that there is no explicit $x^\mu$
dependence in the gauge invariant functions
$g^{I\alpha},\Phi^{i\alpha}$ is used.  Using
$-\d_i\star \mathcal{L}_V=\half\d_i G_I \, F^I$, and the decomposition
$\Phi^{i}|_{{\rm
    der}=0}=\Phi^{i}_0+\Phi^{i}_1+\dots$,
where the $\Phi^{i}_{n}$ depend on undifferentiated scalar
fields and are homogeneous of degree $n$ in $F^I_{\mu\nu}$,
the equation implies that 
\begin{equation}
  \label{eq:28}
  \half M_{IJ}(\phi)\star F^I F^J+\half N_{IJ}(\phi)F^IF^J=0, 
\end{equation}
where 
\begin{equation}
  \label{eq:29}
  M_{IJ}=2\mathcal I_{K(I}{f^K}_{J)}
+\d_i\mathcal
I_{IJ}\Phi^{i}_0, 
\end{equation}
\begin{equation}
  \label{eq:27}
  \quad  N_{IJ}=2\mathcal R_{K(I}{f^K}_{J)}+2h_{IJ}+\d_i\mathcal
  R_{IJ}\Phi^{i}_0,
\end{equation}
by using that $h_{IJ}=h_{JI}$ on account of \eqref{Y1}.
When taking an Euler-Lagrange derivative of \eqref{eq:28} with respect
to $A_\mu^I$, one concludes that both terms have to vanish separately, 
\begin{equation}
M_{IJ}=0,\quad N_{IJ}=0.\label{eq:19}
\end{equation}
Setting $\phi^i=0$ and using \eqref{eq:33} then gives
\begin{equation}
  \label{eq:67}
  f^{(\mathcal I(0))}_{IJ}+f^{(\mathcal I(0))}_{JI}=-(\d_i\mathcal
  I_{IJ})(0)\Phi^i_0(0),\quad 
 2h_{IJ}=-(\d_i\mathcal R_{IJ})(0)\Phi^i_0(0).
\end{equation}
where the abelian index is lowered and raised with $\mathcal I_{IJ}(0)$ and
its inverse. Note that completely skew-symmetric $f^{(\mathcal I(0))}_{IJ}$
solve the equations with $\Phi^i_0(0)=0$, $h_{IJ}=0$. More conditions
are obtained by expanding equations \eqref{eq:19} in terms of power
series in $\phi^i$.

In all examples considered below, the algebra $\mathfrak g_U$ and the
$W$-type symmetries can be entirely determined from the analysis of
this subsection.

\subsection{Electric symmetry algebra}

An important result of our general analysis is that the symmetries of
the action that can lead to consistent gaugings may have a term that
is not gauge invariant. This term is present only in the variation
of the vector potential and is restricted to be linear in the
undifferentiated vector potential, i.e.,
$\delta A^I_{\mu} = f\indices{^I_J} A^J_{\mu} + g^I_\mu$,
$\delta \phi^i = \Phi^i$.  Here $f\indices{^I_J}$ are constants, and
$g^I_\mu$ and $\Phi^i$ are gauge invariant functions.  The symbol
$\delta$ represents the variation of the fields and is of course not
the Koszul-Tate differential.  No confusion should arise as the
context is clear.

It is of interest to investigate a subalgebra of the gaugeable
symmetries, obtained by restricting oneself from the outset to
transformations of the gauge potentials that are linear and
homogeneous in the undifferentiated potentials and to transformations
of the scalars that depend on undifferentiated scalars alone,
\begin{equation}
  \label{eq:64}
\delta A^I_{\mu} = f\indices{^I_J} A^J_{\mu}, \quad
\delta\phi^i = \Phi^i(\phi). 
\end{equation}
This means that one takes $g_\mu^I = 0$ and that the functions
$\Phi^i$ only depend on the undifferentiated scalar fields.  These
symmetries form a sub-algebra $\mathfrak g_e$ that includes the
symmetries usually considered in the supergravity literature and which
is, in this context, called the ``electric symmetry algebra'' (in the
given duality frame) \cite{Trigiante:2016mnt}.  It can be shown to be
a subgroup of the duality group $G \subset Sp(2 n_v, \mathbb{R})$
\cite{Gaillard:1981rj}.  Although our Lagrangians are not necessarily
connected with supergravity, we shall nevertherless call the
symmetries of the form (\ref{eq:64}) ``electric symmetries'' and the
subalgebra $\mathfrak g_e$ the ``electric algebra''.  It need not be a
subalgebra of $Sp(2 n_v, \mathbb{R})$.  It generically does not
exhaust all symmetries and does not contain for example the conformal
symmetries of free electromagnetism.

The transformations of the form (\ref{eq:64}) are symmetries of the
action \eqref{eq:lag} + \eqref{eq:22} if and only if the scalar
variations leave the scalar action invariant separately, and
$f\indices{^I_J},\Phi^i(\phi)$ satisfy
\begin{align}
  \frac{\d \cI}{\d \phi^i} \Phi^i &
= - f^T \mathcal{I} - \mathcal{I} f,\label{var-I}\\
  \frac{\d \cR}{\d \phi^i} \Phi^i &
= - f^T \mathcal{R} - \mathcal{R} f - 2 h, \label{var-R-theta}
\end{align}
where the $h$ are constant symmetric matrices.  In
particular, when the scalar Lagrangian is given by
$\cL_S = \frac{1}{2} g_{ij}(\phi) \d_\mu \phi^i \d^\mu \phi^j$, the first condition means that $\Phi^i$ must be a Killing vector of the
metric $g_{ij}$.    If $U$
and $W$-type symmetries are of electric type, the electric symmetry
algebra contains in addition only $V$-type symmetries of electric
type, i.e., transformations among the undifferentiated scalars alone
that leave invariant both the scalar action and the matrices
$\mathcal I,\mathcal R$ (i.e., that satisfy $\delta S_S=0$ and
\eqref{var-I}, \eqref{var-R-theta} with $0$'s on the right hand
sides). This will be the case in all examples below. In particular,
the $f$'s, and thus also the gauge algebra, will be the same for
$\mathfrak g_U$ and $\mathfrak g_e$. The $h$ matrix is determined by the transformation parameters $\Phi^i$ and the parity-odd term $\mathcal{R}$ of the action via (\ref{var-R-theta}).

We then suppose that we have a basis of
symmetries of the action of this form,
\begin{align}
\delta_\Gamma A^I_{\mu} &= (f_{\Gamma})\indices{^I_J}
                          A^J_{\mu}, \label{global-sym-A}\\ 
\delta_\Gamma \phi^i &= \Phi_{\Gamma}^i(\phi). \label{global-sym-phi}
\end{align}
When compared to the previous sections, the index $\Gamma$ can take
$u$, $v$ or $w$ values. Only the $f_u$ matrices are non-zero. The
$h_\Gamma$ matrices are non-vanishing only for $\Gamma = u$ or
$w$. When $h_\Gamma \neq 0$, the Lagrangian is only invariant up to a
total derivative.

Closure of symmetries of this form then implies
\begin{align}
[f_\Delta, f_\Gamma] &= -C\indices{^\Sigma_{\Delta\Gamma}} f_\Sigma , \label{X-alg}\\
f_\Gamma^T h_\Delta - f_\Delta^T h_\Gamma + h_\Delta f_\Gamma -
  h_\Gamma f_\Delta &
= - C\indices{^\Sigma_{\Delta\Gamma}} \,h_\Sigma , \label{5idGlobal} \\
  \frac{\partial \Phi^i_\Delta}{\partial \phi^j} \, \Phi^j_\Gamma
  - \frac{\partial \Phi^i_\Gamma}{\partial \phi^j} \, \Phi^j_\Delta &
= -C\indices{^\Sigma_{\Delta\Gamma}} \Phi^i_\Sigma . \label{F-alg}
\end{align}
Decomposing the indices into $U$, $W$ and $V$ type, this is consistent
with the relations of section \ref{sec:globalsymmetries} with $\lambda_u^w=0$.
Let us note that (\ref{var-R-theta}) expresses the surface term in the variation of the action. Equation (\ref{5idGlobal}) follows by commutation.

\subsection{Restricted first order deformations}

We now limit ourselves to first order deformations of the master
action with the condition that all infinitesimal gaugings come from
symmetries that belong to the electric symmetry algebra above. In
order to simplify formulas, we will no longer make the distinction
between $U$-, $W$- and $V$-type which can easily be recovered.

According to section \ref{sec:parametrization}, the deformations are
parametrized through electric symmetries by a matrix $k^\Gamma_I$,
with
\begin{align}
f\indices{^I_{JK}} &= (f_\Gamma)\indices{^I_J} k^\Gamma_K , \label{eq:f-k} \\
\Phi^i_I(\phi) &= \Phi^i_\Gamma(\phi) k^\Gamma_I , \label{eq:phi-k} \\
X_{IJ,K} &= (h_\Gamma)_{IJ} k^\Gamma_K .
\end{align}
The linear constraints \eqref{linf} -- \eqref{465} on the matrix
$k^\Gamma_K$ become
\begin{align}
(f_\Gamma)\indices{^I_J} k^\Gamma_K + (f_\Gamma)\indices{^I_K}
  k^\Gamma_J &= 0 ,
\label{eq:constr-antisymmetry}\\
h_{\Gamma \, (IJ} k^\Gamma_{K)} &= 0. \label{eq:constr-chernsimons}
\end{align}
They guarantee that the first order deformation of the master action
is given by
\begin{equation}
S^{(1)} = \int \!d^4\!x\,\left( a_2 + a_1 + a_0 \right),
\end{equation}
where 
\begin{equation}
  a_2 = \frac{1}{2} C^*_I f\indices{^I_{JK}} C^J C^K
\end{equation}
encodes the first order deformation of the gauge algebra and
\begin{equation}
  a_1 = A^{*\mu}_I f\indices{^I_{JK}} A^J_\mu C^K + \phi^*_i \Phi^i_K C^K
\end{equation}
encodes the first order deformation of the gauge symmetries.  When
taking \eqref{eq:f-k} and \eqref{eq:phi-k} into account, this
deformation of the gauge symmetries corresponds to gauging the
underlying global symmetries by using local parameters
$\eta^\Gamma (x) = k^\Gamma_I \epsilon^I (x)$. The deformation $a_0$
of the Lagrangian is given by the sum of three terms:
\begin{align}
  a^\text{(YM)}_0 &= \frac{1}{2} (\star G_I)^{ \mu\nu}
 f\indices{^I_{JK}} A^J_\mu A^K_\nu , \\
a^\text{(CD)}_0 &= J^\mu_K A^K_\mu ,  \\
  a^\text{(CS)}_0 &= \frac{1}{3} X_{IJ,K} \epsilon^{\mu\nu\rho\sigma}
F^I_{\mu\nu} A^J_\rho A^K_\sigma.
\end{align}
The terms $a^\text{(YM)}$ and $a^\text{(CD)}$ are exactly those
necessary to complete the abelian field strengths and ordinary
derivatives of the scalars into covariant quantities. The term $a^\text{(CD)}$ is responsible for charging the matter fields. The Chern-Simons
term $a_0^\text{(CS)}$ appears when $h_\Gamma \neq 0$: its role is to
cancel the variation
\begin{equation} \label{eq:CSobstr}
  \delta \mathcal{L} = - \frac{1}{4} \eta^\Gamma h_{\Gamma \,IJ}\,
  \varepsilon^{\mu\nu\rho\sigma} F^I_{\mu\nu} F^J_{\rho\sigma}
\end{equation}
that is no longer a total derivative when $\eta^\Gamma = k^\Gamma_I
\epsilon^I (x)$ \cite{deWit:1984rvr,deWit:1987ph}.

\subsection{Complete restricted deformations}

The second order deformation $S^{(2)}$ to the master action is then
determined by the first order deformation through equation
\eqref{eq:43}. As discussed in section
\ref{sec:parametrization-2nd-order}, the existence of $S^{(2)}$
imposes additional quadratic constraints on the matrix $k^\Gamma_I$,
\begin{equation} \label{eq:quadraticconstraint}
  k\indices{^{\Gamma}_I} k\indices{^{\Delta}_J}
  C\indices{^{\Sigma}_{\Gamma\Delta}} -
  (f_{\Gamma})\indices{^K_I} k\indices{^\Gamma_J}
  k\indices{^{\Sigma}_K} = 0.
\end{equation}

Explicit computation shows that $S^{(2)}$ can be chosen such that
there is no further deformation of the gauge symmetries or of their
algebra. The second order terms in the Lagrangian are exactly those
necessary to complete abelian field strengths
$F^I_{\mu\nu} = \partial_\mu A^I_\nu - \partial_\nu A^I_\mu$ and
ordinary derivatives of the scalars to non-abelian field strengths and
covariant derivatives,
\begin{align}
  \mathcal{F}^I_{\mu\nu} &= \partial_\mu A^I_\nu - \partial_\nu A^I_\mu
 + f\indices{^I_{JK}} A^J_\mu A^K_\nu  , \\
D_\mu \phi^i &= \partial_\mu \phi^i - \Phi^i_I(\phi) A^I_\mu .
\end{align}
One also finds a non-abelian completion of the Chern-Simons term $a_0^\text{(CS)}$. 
Putting everything together, the Lagrangian after adding the second order deformation is
\begin{align}\label{Full-Lagrangian}
  \mathcal{L} = \mathcal{L}_S(\phi^i, D_\mu \phi^i) &
  - \tfrac{1}{4}\,\mathcal{I}_{IJ}(\phi)
   \mathcal{F}^I_{\mu\nu} \mathcal{F}^{J\mu\nu} + \tfrac{1}{8}\,
   \mathcal{R}_{IJ}(\phi)\,
    \varepsilon^{\mu\nu\rho\sigma}
 \mathcal{F}^I_{\mu\nu} \mathcal{F}^J_{\rho\sigma} \nonumber \\
    &+ \tfrac{2}{3}\, X_{IJ,K}\, \varepsilon^{\mu\nu\rho\sigma} A^J_\mu
      A^K_\nu \left( \partial_\rho A^I_\sigma + \tfrac{3}{8}\,
      f\indices{^I_{LM}} A^L_\rho A^M_\sigma \right).
\end{align}
The associated action can be checked to be invariant under the gauge
transformations
\begin{align}
  \delta A^I_{\mu} &= \partial_\mu \epsilon^I + f\indices{^I_{JK}}
 A^J_{\mu} \epsilon^K , \label{gauge-sym-A}\\
  \delta \phi^i &= \epsilon^I \Phi_I^i(\phi). \label{gauge-sym-phi}
\end{align}
This is equivalent to the fact that the deformation stops at second
order, i.e, that $S=S^{(0)} + S^{(1)} + S^{(2)}$ gives a solution to
the master equation $(S,S)=0$.

Checking directly the invariance of this action under
\eqref{gauge-sym-A} -- \eqref{gauge-sym-phi} without first
parametrizing $f\indices{^I_{JK}}$, $\Phi_I^i(\phi)$ and $X_{IJK}$
through symmetries requires the use of the linear identities
\begin{equation}
f\indices{^I_{JK}} = f\indices{^I_{[JK]}}, \quad X_{(IJ,K)} = 0
\end{equation}
and of the quadratic ones
\begin{align}
f\indices{^I_{J[K_1}}f\indices{^J_{K_2K_3]}} &= 0, \\
  f\indices{^K_{I[L}} X_{M]J,K} + f\indices{^K_{J[L}} X_{M]I,K}
  - \tfrac{1}{2}\, X_{IJ,K} f\indices{^K_{LM}} &= 0, \\
  \frac{\d \Phi^i_I}{\d \phi^j} \Phi^j_J -
  \frac{\d \Phi^i_J}{\d \phi^j} \Phi^j_I + f\indices{^K_{IJ}} \Phi^i_K &= 0.
\end{align}
In terms of $k^\Gamma_I$, these three quadratic identities all come
from the single quadratic constraint \eqref{eq:quadraticconstraint}
once the algebra of global symmetries \eqref{X-alg} -- \eqref{F-alg}
is taken into account.

\subsection{Remarks on $GL(n_v)$ transformations}
\label{sec:frame-transf}

Consider a linear field redefinition of the
abelian vector potentials, $A^I_\mu= {M^I}_J{A'}^J_\mu$ with
$M\in GL(n_v)$. Such a transformation gives rise to a trivial
infinitesimal gauging which corresponds to the antifield
independent part of the trivial ghost number $0$ cocycle
\begin{equation}
S^{(1)}_{triv.} = (S^{(0)},\Xi_s),\quad \Xi_s={f_s^I}_J[d^4x\, C^*_I C^J+\star A^*_I
A^J],\label{eq:46}
\end{equation}
with $f_s\in \mathfrak{gl}(n_v,\mathbb R)$.

Two remarks are in order.

The first concerns the relation to the algebra $\mathfrak{g}_U$
defined in section \ref{sec:globalsymmetries}. It can also be defined
as the largest sub-algebra of $\mathfrak{gl}(n_v,\mathbb R)$ that can
be turned into symmetries of the theory by adding suitable gauge
invariant transformations of the vector and scalar fields, or in other
words, for which there exists a gauge invariant $K_u$ of ghost number
$-1$ such that $(S^{(0)},\Xi_u+K_u)=0$.

In particular, \eqref{eq:352} and \eqref{linf} for $g=0$, as well as
\eqref{eq:quadu}, can be summarized as follows: non-trivial $U$-type
gaugings require the existence of a map (described by $k^u_K$) from
the defining representation of the symmetry algebra
$\mathfrak g_U\subset \mathfrak{gl}(n_v,\mathbb R)$ into the adjoint
representation of the $n_v$-dimensional gauge algebra $\mathfrak g_g$.
  
The second remark is about families of Lagrangians related by linear
transformations of the vector potentials among themselves. It is
sometimes useful not to work with fixed (canonical) values for various
$GL(n_v)$ tensors that appear in the action. Instead, one considers
the deformation problem for sets of Lagrangians parametrized by
arbitrary $GL(n_v)$ tensors, for instance generic non-degenerate
symmetric $\mathcal I_{MN}$ and symmetric $\mathcal R_{MN}$ that
vanish at the origin of the scalar field space.

If the tensors of two such Lagrangians are related by a $GL(n_v)$
transformation, they should be considered as equivalent. Indeed, the
local BRST cohomology for all members of such an equivalence class are
isomorphic and related by the above anti-canonical field
redefinition. In particular, all members of the same equivalence class
have isomorphic gaugings.

All general considerations and results on local BRST cohomology above
apply in a unified way to all equivalence classes. When one explicitly
solves the obstruction equation \eqref{eq:16b} (for instance at $g=-1$
in order to determine the symmetries), the results on local BRST
cohomologies do depend on the various equivalence classes.

\subsection{Comparison with the embedding tensor constraints}

In the embedding tensor formalism
\cite{deWit:2002vt,deWit:2005ub,deWit:2007kvg,Samtleben:2008pe,%
  Trigiante:2016mnt}\footnote{See also \cite{Coomans:2010xd}, where a relation between the embedding tensor formalism and the BRST-BV antifield formalism has been considered with a different purpose.}, the possible gaugings are described by the
embedding tensor
$\Theta\indices{_M^\alpha}=(\Th\indices{_I^\a},\Th^{I\a})$ with
electric and magnetic components, which satisfies a number of linear
and quadratic constraints. In this notation, the index $I$ runs from
$1$ to $n_v$, while $\a$ runs from $1$ to the dimension of the group
$G$ of invariances of the equations of motion of the initial
Lagrangian \eqref{eq:lag}. More precisely, $G$ is defined only as
the group of transformations that act linearly on the field strengths
$F^I$ and their ``magnetic duals'' $G_I$, and whose action on the
scalars contains no derivatives. This coincides with the group of
symmetries of the first order Lagrangian discussed in
\cite{Henneaux:2017kbx} which are of the restricted form
\eqref{eq:64}.

As explained in section 3 of \cite{Trigiante:2016mnt}, one can always
go to a duality frame in which the magnetic components of the
embedding tensor vanishes, $\Th^{I\a} = 0$. Moreover, only the
components $\Th\indices{_I^\Gamma}$ survive, where $\Gamma$ runs over
the generators of the electric subgroup $G_e \subset G$ that act as
local symmetries of the Lagrangian in that frame. Then, the gauged
Lagrangian in the electric frame is exactly the Lagrangian
\eqref{Full-Lagrangian}, where the matrix $k$ is identified with the
remaining electric components of the embedding tensor,
$k^\Gamma_I = \Th\indices{_I^\Gamma}$ (or
$\Th\indices{_{\hat{I}}^\Gamma}$ in the notation of
\cite{Trigiante:2016mnt}). The linear and quadratic constraints on the
embedding tensor then agree with the constraints on $k$. More
precisely, the constraints (3.11), (3.12) and (3.39) of
\cite{Trigiante:2016mnt} in the electric frame correspond to our
\eqref{eq:quadraticconstraint}, \eqref{eq:constr-antisymmetry} and
\eqref{eq:constr-chernsimons} respectively. As explained in sections
\ref{sec:parametrization} and \ref{sec:parametrization-2nd-order}, the
constraints can be refined using the split corresponding to the
various ($U$, $W$, $V$) types of symmetry.

It was shown in \cite{Henneaux:2017kbx} that the embedding tensor
formalism does not allow for more general deformations than those of
the Lagrangian \eqref{eq:lag} studied in this paper. Indeed, their
BRST cohomologies are isomorphic even though the field content and
gauge transformations are different.  Conversely, as long as one
restricts the attention to the symmetries of \eqref{eq:lag} that are
of the electric type \eqref{eq:64}, we showed that the embedding
tensor formalism captures all consistent deformations that deform the
gauge transformations of the fields.

\section{Applications}
\label{sec:applications}

\subsection{Abelian gauge fields: $U$-type gauging}
\label{sec:481}

As a first example, let us consider the case where we have no scalars,
$\mathcal{I}_{IJ} = \delta_{IJ}$, $\mathcal{R}_{IJ} = 0$. The
Lagrangian is then simply
\begin{equation}
  \label{eq:YM} \mathcal{L} = -\frac{1}{4}
  \delta_{IJ} F^I_{\mu\nu} F^{J\mu\nu}.
\end{equation}

From \eqref{eq:67}, it can be shown that $U$-type symmetries are of
electric form. Furthermore, there are no $W$-type
symmetries. We have in this case
$\mathfrak{g}_U = \mathfrak g_e=\mathfrak{so}(n_v)$. 

The vector fields transform in the fundamental representation of
$\mathfrak{so}(n_v)$. A basis of the Lie algebra $\mathfrak{so}(n_v)$ may be labeled by an
antisymmetric pair of indices $[LM]$ that now plays the role of the
index $u$,
\begin{equation}
\delta_{[LM]} A^I_\mu = (f_{[LM]})\indices{^I_J} A^J_\mu,\quad 
\qquad (f_{[LM]})\indices{^I_J} = \half(\delta^I_L \delta_{JM} - \delta^I_M
                                \delta_{JL})\label{eq:so(n)}. 
\end{equation}

Concerning the associated gaugings, the matrices
${f_{[LM]}}_{IJ}=\delta_{II'}(f_{[LM]})\indices{^{I'}_J}$ are
antisymmetric in $I,J$; therefore, the structure constants of the
gauge group
\begin{equation}
f^{(\delta)}_{IJK} = (f_{[LM]})\indices{_{IJ}}  k^{LM}_K
\end{equation}
are automatically antisymmetric in their first two indices. The
constraints on $k^{LM}_K$ ensure antisymmetry in the last two indices
(which in turn implies total antisymmetry) and the Jacobi
identity. Moreover, any set of totally antisymmetric structure
constants can be obtained in this way by taking
$k^{LM}_K =f\indices{^{LM}_K}$, as can be easily seen
using the expression for $f_{[LM]}$ given above.

We thereby recover the result of \cite{Barnich:1993pa,Ramond:1986vt} stating
that the most general deformation of the free Lagrangian \eqref{eq:YM}
that is not of $V$ or $I$-type is given by the Yang-Mills Lagrangian
with a compact gauge group of dimension equal to the number of vector fields.

{\bf Remark:} Note that Poincar\'e (conformal) symmetries (for $n=4$)
are of $V$-type if one allows for $x^\mu$-dependent local
functions. If such a dependence is allowed for $U,W$-type symmetries
and gaugings as well, results can be very different. For instance, as
shown by equation (13.21) of\cite{Barnich:2000zw}, if $n\neq 4$, there
are additional $U$-type symmetries described by the cohomology class
\begin{equation}
  \label{eq:36}
U^{-1}=  d^nx f_{(IJ)}\big[C^{*I}C^J+A^{*\mu I} A_\mu^J+\frac{2}{n-4}F^I_{\mu\nu}x^\mu
  A^{*\nu J}\big],
\end{equation}
where indices $I,J,\dots$ are raised and lowered with the Kronecker
delta. The associated Noether current can be obtained by working out
the descent equation following \eqref{eq:18b},
$sU^{-1}+d (f_{(IJ)}[\star A^{*I}C^J+\star F^IA^J+{J_U}^{IJ}] ) = 0$, where 
\begin{equation}
  \label{eq:58}
  {J_U}^{IJ}=\frac{2}{n-4}(T_{\mu\nu})^{IJ} x^\nu\star dx^\mu,\quad
  ({T^\mu}_\nu)^{IJ}=F^{(I\vert \mu\rho}F^{\vert J)}_{\rho\nu}+\frac{1}{4}
  F^{I\alpha\beta}F^J_{\alpha\beta}\delta^\mu_\nu.
\end{equation}

In other words, $\mathfrak g_U=\mathfrak{gl}(n_v)$. Note also
that these $U$-type symmetries involve a non-vanishing ${}^U
g^I_\mu$. It has furthermore been shown in section 13.2.2
of\cite{Barnich:2000zw} that there are associated $U$-type gaugings
and cohomology classes in higher ghost numbers. In the present
context, they are obtained as follows: the role of $u$ for the
additional symmetries is played by a symmetric pair of indices $(LM)$,
\begin{equation}
  \label{eq:37}
  \delta_{(LM)} A^I_\mu= (f_{(LM)})\indices{^I_J} A^J_\mu,\quad
  (f_{(LM)})\indices{^I_J}
  =\half(\delta^I_L \delta_{JM} + \delta^I_M
                                \delta_{JL}). 
\end{equation}
Once the linear constraints \eqref{linf} on
$k^{(LM)}_{K_1\dots K_{g+1}}$ are fulfilled, the associated $U$-type
gaugings and higher cohomology classes can be read off from equation
\eqref{eq:18b} when taking \eqref{eq:352}.

After multiplying \eqref{eq:36} by $n-4$, it represents for $n=4$ the
$V$-type symmetry associated with the dilatation of the conformal
group. The associated cubic and higher order vertices for the full
conformal group have been studied in detail in \cite{Brandt:2001hs}.

\subsection{Abelian gauge fields with uncoupled scalars:
  $U,V$-type gaugings}

We now take the case
\begin{equation}
  \mathcal{L} = \mathcal{L}_S(\phi^i, \partial_\mu \phi^i)
  - \frac{1}{4} \delta_{IJ} F^I_{\mu\nu} F^{J\mu\nu},
\end{equation}
where there is no interaction between the scalars and the vector
fields. The $\mathfrak{g}_U$ algebra is again $\mathfrak{so}(n_v)$
and there are no $W$-type symmetries.

The electric symmetry algebra is the direct sum of
$\mathfrak{so}(n_v)$ with the electric $V$-type symmetry algebra
$\mathfrak g_s$ of the scalar Lagrangian. The matrices $f_\Gamma$
split into two groups and are given by
\begin{equation}
  (f_\alpha)\indices{^I_J} = 0, \qquad (f_{[LM]})\indices{^I_J}
  = \half(\delta^I_L \delta_{JM} - \delta^I_M \delta_{JL})
\end{equation}
where $\alpha = 1, \dotsc, \dim \mathfrak g_s$ labels the $G_s$
generators and the antisymmetric pair $[LM]$ labels the $SO(n_v)$
generators as before. The matrix $k^\Gamma_I$ accordingly splits in
two components $k^{LM}_I$ and $k^\alpha_I$. The constraints on
$k^{LM}_I$ again amount to the fact that the quantities
$f\indices{^I_{JK}} = (f_{[LM]})\indices{^I_J} k^{LM}_K$ are the
structure constants of a compact Lie group. The constraint on
$k^\alpha_I$ tells us that the gauge variations
\begin{equation}
\delta \phi^i = \epsilon^I k^\alpha_I \Phi^i_\alpha(\phi)
\end{equation}
close according to the structure constants $f\indices{^I_{JK}}$. In
the case where these variations are linear,
$\Phi^i_\alpha(\phi) = (t_\alpha)\indices{^i_j} \phi^j$, the constraint
is that the matrices $T_I = k^\alpha_I t_\alpha$ form a representation
of the gauge group, $[T_I, T_J] = f\indices{^K_{IJ}} T_K$.

\subsection{Bosonic sector of $\mathcal{N}=4$ supergravity}
\label{sec:N=4}
Neglecting gravity, the bosonic sector of $\mathcal{N}=4$ supergravity
is given by two scalar fields parametrizing the coset $SL(2,\R)/SO(2)$
along with $n_v=6$ vector fields
{\cite{Cremmer:1977tc,Das:1977uy,Cremmer:1977tt,Cremmer:1979up}\footnote{For generalisations
    of the models treated in this section, see \cite{Tsokur:1994qt}
    for the couplings of $\mathcal{N}=4$ supergravity to an
    arbitrary number of vector supermultiplets.}.  We study three formulations of this model, where we
determine the symmetry algebras $\mathfrak{g}_U$ and $\mathfrak{g}_e$
and the allowed gaugings. In all formulations, the scalar Lagrangian
is determined by
\begin{equation}
\phi^i=(\phi,\chi),\quad g_{ij}={\rm diag}(1,e^{2\phi}),\quad
V=0\label{eq:14a}. 
\end{equation}
They differ by the form of the matrices $\mathcal{I}$ and $\mathcal{R}$.

\subsubsection*{$SO(6)$ formulation}

The vector Lagrangian
  \eqref{eq:lag} is determined by  
\begin{equation}
  \mathcal{I}_{IJ} = e^{-\phi}\delta_{IJ},\quad \mathcal{R}_{IJ}=
  \chi\delta_{IJ}.\label{eq:10a} 
\end{equation}
When $f^{(\delta)}_{IJ}$ is antisymmetric, the transformations
$\delta A_\mu^I={f^I}_J A^J_\mu$ define an $\mathfrak{so}(6)$
sub-algebra of $U$-type symmetries on their own. Note also that we can
assume $h_{IJ}$ to be symmetric. Equations \eqref{eq:67} then imply
that the traceless parts of $f^{(\delta)}_{IJ},h_{IJ}$ have to vanish.

If $f^{(\delta)}_{IJ}=\delta_{IJ}\eta^0$, equations \eqref{eq:67} are
solved with $h_{IJ}=0$, $\Phi^\phi_0(0)=2\eta^0, \Phi^\chi_0(0)=0$. It
then follows that \eqref{eq:19} are solved with
\begin{equation}
\Phi_0^\phi=2\eta^0,\quad
\Phi_0^\chi=-2\eta^0\chi\label{eq:41}.
\end{equation}
Equation \eqref{eq:17a} is
then also solved with $g^I=0$ and
$I^3=2\eta^0(e^{2\phi}\star d\chi\chi-\star d\phi)$. According to
equation \eqref{eq:18b}, the associated cohomology class is given by
\begin{equation}
    \label{eq:15a} \omega^{-1,4}= \eta^0[d^4x C^{*}_IC^I+\star A^{*}_IA^I
+2(\star\phi^*-\star\chi^*\chi)],
\end{equation}
with $s\omega^{-1,4}+d[\eta^0(\star A^{*}_IC^I+G_IA^I)+I^3]=0$. This
cohomology class encodes the symmetry
\begin{equation}
  \delta A^I_\mu = \eta^0 A^I_\mu, \quad \delta \phi=2 \eta^0, \quad \delta\chi=-2\eta^0 \chi,
\end{equation}
with
$\delta \mathcal L_0=0$. The associated Noether current is given by 
\begin{equation}
  j^\mu=[-(e^{-\phi}F^{\mu\lambda}_I-\half\chi
\epsilon^{\mu\lambda\rho\sigma}F_{I\rho\sigma})A_\lambda^I-2\d^\mu\phi+2\chi
e^{2\phi}\d^\mu\chi].
\end{equation}
It cannot be made gauge invariant through allowed redefinitions. 

For $f_{IJ}=0$, $h_{IJ}=\eta^+\delta_{IJ}$,
$\Phi^\chi_0=-2\eta^+,\Phi^\phi_0=0$ is a solution to the full problem
\eqref{eq:17a} since
$\half F^I F_I=s\star \chi^*+d(e^{2\phi}\star d\chi)$. This gives then
the only class of $W$-type, which is also of restricted type. More
explicitly, $W^{-1} = \star\chi^*$ with
$s\star \chi^*+d(-\half A^IF_I+ e^{2\phi}\star d\chi)=0$. The symmetry
it describes is $\delta \chi= \eta^+$ with the associated Noether
current given above that can again not be made gauge invariant.

The algebra $\mathfrak{g}_U$ is therefore isomorphic to
$\mathfrak{so}(6)\oplus \mathfrak h$, where $\mathfrak h$ is the
sub-algebra of $\mathfrak{sl}(2,\mathbb R)$ generated by diagonal
traceless matrices. It is a sub-algebra of the electric symmetry
algebra $\mathfrak{g}_e = \mathfrak{so}(6)\oplus \mathfrak b^+$, where
$\mathfrak b^+$ corresponds to the sub-algebra of
$\mathfrak{sl}(2,\mathbb R)$ of upper triangular matrices. The
electric algebra acts as
\begin{equation}
\delta \phi = 2 \h^0, \quad
\delta \chi = - 2 \h^0 \chi + \h^+,\quad
  \delta A^I_\mu = \h^0 A^I_\mu + \h^{LM} (f_{[LM]})
            \indices{^I_J} A^J_\mu \label{eq:N=4deltaA}.
\end{equation}
Accordingly $f_\Gamma=(f_0,f_+,f_{[LM]})$ where
$f_{[LM]}$ are given in \eqref{eq:so(n)}, while  
\begin{equation}
(f_0)\indices{^I_J} = \delta^I_J, \qquad (f_+)\indices{^I_J} = 0.
\end{equation}
The matrix $\mathcal{R}_{IJ} = \chi \delta_{IJ}$ transforms as
\begin{equation}
\delta \mathcal{R}_{IJ} = - 2 \h^0 \mathcal{R}_{IJ} + \h^+ \delta_{IJ}.
\end{equation}
Therefore, contrary to the previous examples, the tensor
$h_{\Gamma IJ}$ has a non-vanishing component $h_{+IJ} = - \frac{1}{2}
\delta_{IJ}$.

The generalized structure constants
$f\indices{^I_{JK_1\dots K_{g+1}}} = (f_\Gamma)\indices{^I_J}
k^\Gamma_{K_1\dots K_{g+1}}$ are then
\begin{equation}
f\indices{^I_{JK_1\dots K_{g+1}}} = \delta^I_J k^0_{K_1\dots K_{g+1}} + \half
(f_{[LM]})\indices{^I_J}  k^{LM}_{K_1\dots K_{g+1}} .
\end{equation}
The linear constraint \eqref{linf} now implies that
$k^0_{K_1\dots K_{g+1}}=0$ as can be seen by taking the first three indices
equal and using the antisymmetry of the matrices $f_{[LM]}$, while
$k^{(\delta)}_{LM K_1\dots K_{g+1}}$ is restricted to be completed
skew-symmetric in all indices. In the same way, the linear constraint
\eqref{465} implies that $k^+_{K_1\dots K_{g+1}}=0$. Indeed, it reduces to
\begin{equation}
\delta_{(IJ} k^+_{K_1)\dots K_{g+1}} = 0,
\end{equation}
from which we deduce $k^+_{K_1\dots K_{g+1}} = 0$ by taking the first three indices
equal.

It follows that
there are no cohomology classes of $W$-type when $g\geqslant 0$ and that
the only cohomology classes of $U$-type when $g\geqslant 0$ are given by
\begin{equation}
    \label{eq:11} [d^4x C^{*I}\d_I+\star A^{*I}A^J\d_J\d_I+\half
    G^I A^J A^K\d_K\d_J\d_I]\Theta,
  \end{equation}
with $\Theta$ a polynomial in $C^I$ of ghost number $\geqslant 1$.

In particular, the symmetries of $\mathfrak b^+$ cannot be gauged and
the gauge algebra is given by a compact sub-algebra of
$\mathfrak{so}(6)$. The gauged Lagrangian is the original one,
except that the abelian field strengths are replaced by non-abelian
ones.

\subsubsection*{Dual $SO(6)$ formulation}
We now have
\begin{equation}
    \mathcal{I}_{IJ} = \frac{1}{e^{-\phi}+\chi^2e^\phi}\delta_{IJ},\quad 
    \mathcal{R}_{IJ}= -\frac{\chi
      e^\phi}{e^{-\phi}+\chi^2e^\phi}\delta_{IJ},\label{eq:18}
  \end{equation}
and the same analysis gives similar conclusions:
\begin{enumerate}

\item We still have the cohomology classes \eqref{eq:11} since we
  still have that $\mathcal I_{IJ}$ and $\mathcal R_{IJ}$ are proportional to $\delta_{IJ}$ .
  
\item There are no additional gaugings or cohomology classes in ghost
  number higher than $0$ of $U$ or $W$-type.

\item The only additional
  non-covariantizable characteristic cohomology comes from two
  additional solutions to \eqref{eq:67}.
\end{enumerate}
The first of these additional solutions is of electric
  $U$-type and comes from
  $f^{(\delta)}_{IJ}=\tilde \eta^0\delta_{IJ}$, $h_{IJ}=0$, with
  $\Phi^\phi_0(0)=-2\tilde \eta^0$, $\Phi^\chi_0(0)=0$. Equation \eqref{eq:19}
  reduces to
  \begin{equation}
    \label{eq:39}
    2 \tilde \eta^0 \mathcal I_{IJ}+\d_i \mathcal I_{IJ}\Phi^i_0=0,\quad
    2 \tilde \eta^0
    \mathcal R_{IJ}+2h_{IJ}\delta_{IJ}+\d_i  \mathcal
    R_{IJ}\Phi^i_0=0, 
  \end{equation}
  and is solved by
  \begin{equation}
    \label{eq:9}
    h_{IJ}=0\quad \Phi^\phi_0=-2\tilde \eta^0,\quad  
    \Phi^\chi_0=2\tilde \eta^0\chi.
  \end{equation}
This gives also a solution to the full problem since this
transformation leaves the scalar field Lagrangian invariant.
According to equation \eqref{eq:18b}, the associated cohomology class is given by
\begin{equation}
    \label{eq:15b} \omega^{-1,4}= \tilde \eta^0[d^4x C^{*}_IC^I+\star A^{*}_IA^I-2(\star\phi^*-
\star\chi^*\chi)].
\end{equation} 
The second is of restricted $W$-type and comes from
$f^{(\delta)}_{IJ}=0$ while $h_{IJ}=\tilde \eta^+\delta_{IJ}$ with
$\Phi^\phi_0(0)=0$, $\Phi^\chi_0(0)=2\tilde \eta^+$. Equation
\eqref{eq:19} reduces to
\begin{equation}
  \label{eq:40}
  \d_i \mathcal I_{IJ}\Phi^i_0=0,\quad 2\tilde \eta^+\delta_{IJ}+\d_i  \mathcal
    R_{IJ}\Phi^i_0=0,
\end{equation}
and is solved by
\begin{equation}
\Phi^\chi_0=2\tilde \eta^+(e^{-2\phi}-\chi^2),\quad 
\Phi^\phi_0=4\tilde \eta^+\chi\label{eq:38},
\end{equation}
This is also a solution to the full problem since these
transformations leave the scalar field Lagrangian invariant. 
The associated cohomology class is given by
\begin{equation}
  \label{eq:12a}
  2\tilde \eta^+ [\star \phi^*2\chi+\star\chi^*(e^{-2\phi}-\chi^2)]. 
\end{equation}
In this case, we therefore have
$\mathfrak{g}_U = \mathfrak{so}(6)\oplus \mathfrak h \subset
\mathfrak{g}_e = \mathfrak{so}(6)\oplus \mathfrak{b}^-$, where
$\mathfrak b^-$ is the sub-algebra of $\mathfrak{sl}(2,\mathbb R)$ of
lower triangular matrices.

Again, the symmetries of $\mathfrak b^+$ cannot be gauged and the gauge algebra is given by a compact sub-algebra of $\mathfrak{so}(6)$.

\subsubsection*{$SO(3)\times SO(3)$ formulation}
The indices split as $I=(A,A')$, where $A,A'=1,2,3$, and we have 
  \begin{equation}
\begin{split}
    \mathcal
    I_{IJ}={\rm diag}(\mathcal I_{AB},\mathcal I_{A'B'}),\quad
    \mathcal R_{IJ}={\rm diag}( \mathcal R_{AB},\mathcal R_{A'B'})\label{eq:17},\\
\mathcal I_{AB}=e^{-\phi}\delta_{AB},\quad \mathcal
I_{A'B'}=\frac{1}{e^{-\phi}+\chi^2e^\phi}\delta_{A'B'},\\
\mathcal R_{AB}=\chi\delta_{AB},\quad \mathcal R_{A'B'}=-\frac{\chi
      e^\phi}{e^{-\phi}+\chi^2e^\phi}\delta_{A'B'}.
  \end{split}
\end{equation}
Spelling out equation \eqref{eq:19} gives 
\begin{equation}
  \label{eq:21}
  \begin{split}
 & \mathcal I_{IL} {f^L}_{J}+\mathcal I_{LJ}{f^L}_{I}+\d_i\mathcal I_{IJ}{\Phi^i_0}=0,\\
& \mathcal R_{IL} {f^L}_{J}+\mathcal R_{JL}{f^L}_{I}+h_{IJ}+h_{JI}+\d_i\mathcal
    R_{IJ}{\Phi^i_0}=0. 
  \end{split}
\end{equation}
Choosing $I=A$, $J=A'$, the first equation reduces to 
\begin{equation}
  \label{eq:30}
  e^{-\phi}f^{(\delta)}_{AA'}+\frac{1}{e^{-\phi}+\chi^2e^\phi}f^{(\delta)}_{A'A}=0.
\end{equation}
Putting $\phi=0=\chi$ and taking the derivative with respect to $\phi$
before putting $\phi=0=\chi$, implies that
$f^{(\delta)}_{AA'}=0=f^{(\delta)}_{A'A}$. When combined with the
linear constraint \eqref{linf}, this implies that the
${f^{I}}_{JK_1\dots K_{g+1}}$'s have to vanish
unless all indices are of $A$, or of $A'$, type respectively, which is
thus a necessary condition to have non trivial $U$-type solutions.

When the $f$'s vanish, the second equation for $I=A$, $J=A'$ gives
$h_{AA'}+h_{A'A}=0$. When combined with the linear constraint
\eqref{465}, this implies that for non-trivial solutions of $W$-type
associated to $X_{IJ,K_1\dots K_{g+1}}$ one again needs all indices to
be either of $A$ or of $A'$ type.

The discussion then reduces to the one we had before in each of the sectors. For $U$-type solutions, this gives in a first stage the
symmetries, gaugings and higher ghost cohomology classes associated with
each of the $SO(3)$ rotations separately. There are again no
additional solutions of $U$ or $W$ type when $g\geqslant 0$.

Only the remaining non-covariantizable symmetries, i.e., solutions of
type $U$ and $W$ at $g=-1$ that correspond to $\mathfrak{b}^\pm$, remain to be discussed. For the $U$ type
solutions, one finds in the first sector that
$f_{AB}=\eta^0\delta_{AB}$ with \eqref{eq:41} holding, while for the
second sector $f_{A'B'}=\tilde \eta^0\delta_{A'B'}$ with (\ref{eq:9})
holding. This gives a solution to the full problem if and only if
$\tilde \eta^0=-\eta^0$. Hence $\mathfrak
g_U=\mathfrak{so}(3)\oplus\mathfrak{so}(3)\oplus \mathfrak h$.
On the other hand the solutions of $W$ type for both sectors are
solutions to the full problem if and only if $\eta^+=\tilde \eta^+=0$ so
that there is no surviving $W$-type symmetry. In particular $\mathfrak
g_e=\mathfrak g_U$.
The symmetry of $\mathfrak{h}$ cannot be gauged, and the gauge algebra is a compact sub-algebra of $\mathfrak{so}(3)\oplus \mathfrak{so}(3)$.

This concludes the discussion with the expected results (see \cite{Das:1977pu,Freedman:1978ra,Gates:1982ct}).

\subsection{Axion models: $W$-type gaugings and anomalies}
\label{subsec:axionmodel}

We now give examples of gaugings and anomalies where the generalized
Chern-Simons term appears. They involve several axions and correspond
to the examples given in \cite{Barnich:2000zw}, equations (12.4) and
(12.6).

The initial Lagrangian is
\begin{equation}
  \mathcal{L}_0 =   
  - \frac{1}{4} F^I_{\mu\nu} F^{J\mu\nu}\,\delta_{IJ}
  - \frac{1}{2}\, \partial_\mu \phi^I \partial^\mu \phi^J\, \delta_{IJ} 
  + \frac{1}{4}\,  \varepsilon^{\mu\nu\rho\sigma}\,  
  F^I_{\mu\nu} F^J_{\rho\sigma} \; \phi_{I}V_{J}\;,
\end{equation}
where $V^I$ is a vector of dimension $n_v$ of unit norm and there are
$n_v$ scalar fields $\phi^I$ whose indices are raised and lowered with
the Kronecker symbol. With respect to the general Lagrangian
\eqref{eq:lag} and \eqref{eq:22}, we have here $g_{ij}=\delta_{IJ}$,
$V=0$, $\mathcal I_{IJ}=\delta_{IJ}$ and
$\mathcal R_{IJ}=2\phi_{(I} V_{J)}$. The example in
\cite{Barnich:2000zw}, equation (12.4), corresponds to the case
$n_v = 2$, with $V_J=\delta^2_J$.

By using equations \eqref{eq:67}, \eqref{eq:27}, one finds
$\mathfrak g_U=\mathfrak{so}(n_v-1)$. The symmetries act like
$\delta_{u} A^{I} = f^{I}{}_{J}\,A^{J}$ and
$\delta_{u} \phi^{I} = f^{I}{}_{J}\,\phi^{J}\,$ for an antisymmetric
symbol $f_{IJ}\,$ that is transverse to the vector $V\,$,
$V_I{f^I}_J=0$. 

To determine the $W$-type symmetries, one needs in particular to solve
equation \eqref{eq:67} with ${f^I}_J=0\,$. This is done through
$\delta \phi^I = \Phi^I =\eta^I$ for independent constants $\eta^I$
with $(h_K)_{IJ} = -\delta_{K(I}V_{J)}\,$. There are thus $n_v$
independent such symmetries. It then follows that a basis of $W$-type
symmetries is given by $\delta_K \phi^I = \delta^I_K$,
$\delta_K A^I_\mu=0\,$, where $K$
plays the role of the index $w$. Furthermore, there are no $V$-type
symmetries of electric type so that
$\mathfrak g_e=\mathfrak{so}(n_v-1)\oplus \mathfrak{u}(1)^{n_v}$.

The linear constraints \eqref{linf} require the
$k^u_{K_1\dots K_{g+1}}$ to be transverse,
$k^u_{K_1\dots K_{g+1}} V^{K_1}=0$. For the associated gaugings, the
rest of the discussion then follows the one in subsection
\ref{sec:481}. For $W$-type cohomology, the linear constraints
\eqref{465} imply that
\begin{align}
V_{(I}\,k_{J K_1)\dots K_{g+1}} = 0\;,\quad k_{I K_1\ldots K_{g+1}} :=
  \delta_{IJ}\,k^J{}_{K_1J_2\ldots K_{g+1}} , 
\end{align}
which is solved if and only if $k$
is a totally antisymmetric rank-$(g+2)$ symbol, $k_{I K_1\ldots
  K_{g+1}}=k_{[I K_1\ldots K_{g+1}]}$. 

For $W$-type gaugings in particular, $g=0$ and one has an
antisymmetric $n_v\times n_v$ matrix. There are no extra quadratic
constraints. The gauge transformations, covariant derivatives and
generalized Chern-Simons term for the deformed theory are then
\begin{equation}
  \begin{split}
    \delta A^I_\mu &= \partial_\mu \epsilon^I, \quad \delta \phi_I =
   -k_{[IJ]} \epsilon^J, \quad D_\mu \phi_I =
   \partial_\mu \phi_I + k_{[IJ]} A^J_\mu, \\
  \cL^\text{(CS)} &= -\tfrac{1}{3} (V_K k_{IJ} + V_I k_{KJ}) \,  
                    \varepsilon^{\mu\nu\rho\sigma} A^I_\mu A^J_\nu 
                    F^K_{\rho\sigma} \;.
  \end{split}
\end{equation}
For a $W$-type anomaly example, let $n_s=n_v=3$ and consider
$V_K=\delta_{K\,3}\,$. The associated anomaly candidate is then
\begin{equation}
    W^{1,4}=d^4x\, \epsilon_{IJK}\,
    \left[
    \tfrac{1}{4}\,\epsilon^{\mu\nu\rho\sigma}
    \,C^I\, A^J_{\mu}\,A^K_{\nu}\, F_{\rho\sigma}^3 
    + C^I\, A^J_{\mu}\,\partial^{\mu}\phi^K 
    -\tfrac{1}{2}\,C^I\,C^J\,\phi^{*K} \right]\ .
\end{equation}
\section{First order manifestly duality-invariant actions}
\label{sec:first-order-actions}

\subsection{Non-minimal version with covariant gauge structure}
\label{sec:non-minimal}

We now investigate the first order formulation \cite{Bunster:2011aw,Henneaux:2017kbx}
of the models discussed previously. Those models are interesting
because they contain more symmetries and therefore potentially more
gaugings. In the original, minimal version, they are given by the
action
\begin{equation} \label{eq:symlag} S = \int \!d^4x \left( \frac{1}{2}
    \Omega_{MN} {B}^{Mi} \dot{{A}}^N_i - \frac{1}{2}
    \mathcal{M}_{MN}(\phi) {B}^M_i {B}^{Ni} \right),
\end{equation}
where the potentials are packed into a vector
\begin{equation}
({A}^M) = (A^I, Z_I), \quad M= 1, \dots, 2n_v,
\end{equation}
and the magnetic fields are
\begin{equation}
{B}^{Mi} = \epsilon^{ijk} \partial_j {A}^M_k .
\end{equation}
The matrices $\Omega$ and $\mathcal{M}(\phi)$ are the
$2n_v \times 2n_v$ matrices
\begin{equation} \label{eq:OMdefa}
\Omega = \begin{pmatrix}
0 & I \\ -I & 0
\end{pmatrix}, \qquad
\mathcal{M} = \begin{pmatrix}
  \mathcal{I} + \mathcal{R}\mathcal{I}^{-1}\mathcal{R} &
  - \mathcal{R} \mathcal{I}^{-1} \\
- \mathcal{I}^{-1} \mathcal{R} & \mathcal{I}^{-1}
\end{pmatrix},
\end{equation}
each block being $n_v\times n_v$. The matrix
$\mathcal{N} = \Omega^{-1} \mathcal{M}$ is symplectic,
$\mathcal{N}^T \Omega \mathcal{N} = \Omega$.

Local BRST cohomology and gaugings for this class of models with
non-covariant gauge symmetries $\delta A^M_i=\partial_i\epsilon^M$
could then be discussed by generalizing the results of
\cite{Bekaert:2001wa} in the presence of coupled scalars.

However, in order to be able to directly use the discussion of local
BRST cohomology developed for the second order covariant Lagrangian in
the case of the first order manifestly duality invariant formulation,
we consider a modification of the non-minimal variant
\cite{Barnich:2007uu} with additional scalar potentials for the
longitudinal parts of electric and magnetic fields. More precisely, we
now take instead of \eqref{eq:symlag} the action
\begin{equation}
  \label{eq:48}
  S[A_\mu^M,D^M,\pi_M,\phi^i]=S_S[\phi]+S_{DP},
\end{equation}
with
\begin{multline}
  \label{eq:49}
  \mathcal L_{DP}=\half [ \Omega_{MN}(\mathcal B^{Mi}+\d^iD^M)(\d_0
  A_i^N-\d_i A_0^N)-\mathcal B^{Mi} \mathcal M_{MN}(\phi)\mathcal
  B^{N}_i ]\\+\pi_M\d_0 D^M -\half \pi_M (\mathcal
  M^{-1})^{MN}\pi_N-\mathcal V(\phi,D).
\end{multline}
Here 
\begin{equation}
\mathcal B^{Mi}=\epsilon^{ijk}\d_j
  A_k^M+\d^i D^M,\label{eq:68}
\end{equation}
spatial indices $i,j,k,\dots$ are raised and lowered with
$\delta_{ij}$ and its inverse, with $\Omega_{MN}$ the symplectic
matrix, 
$\mathcal M_{MN}$ symmetric and invertible and
\begin{equation}
(\d_i\mathcal
V)(0,0)=0=(\d_M\mathcal V)(0,0), \quad \cV (\phi, 0) = 0 . \label{eq:51} 
\end{equation}
The modification with respect to \cite{Barnich:2007uu} consists in the
addition of the kinetic and potential terms for the longitudinal
electric and magnetic potentials in the last line of \eqref{eq:49}.
Defining
\begin{equation}
  \label{eq:53}
  \mathcal F^M_{\mu\nu}=\d_\mu A_\nu^M-\d_\nu A_\mu^M+\star
  S^M_{\mu\nu},\quad \star
  S^M_{0i}=\Delta^{-1}(\Omega^{-1})^{MN}\d_0\d_i\pi_N,\quad \star 
  S^M_{ij}=\epsilon_{ijk}\d^k D^M,
\end{equation}
we have $\mathcal B^{Mi}=\half\epsilon^{ijk}\mathcal F_{jk}$ and can write
\begin{equation}
  \label{eq:54}
  S_{DP}=\int d^4x \frac{1}{4} [\Omega_{MN}\epsilon^{ijk} (\mathcal
  F^M_{jk}+\star S^M_{jk})\mathcal F^N_{0i}-\mathcal F^M_{ij} \mathcal
  M_{MN}\mathcal F^{Nij}-2  \pi_M (\mathcal
  M^{-1})^{MN}\pi_N-4\mathcal V],  
\end{equation}
where a total derivative has been dropped.

The gauge invariances are then doubled but still of the same covariant
form as in the second order Lagrangian case,
\begin{equation}
  \label{eq:52}
  \delta A^M_\mu=\d_\mu\epsilon^M,\quad \delta D^M=0,\quad
  \delta\pi_M=0,\quad \delta \phi^i=0.
\end{equation}
The equations of motion for the gauge and scalar potentials are
determined by the vanishing of
\begin{equation}
  \label{eq:50a}\begin{split}
 &   \vddl{\mathcal L_{DP}}{\pi_M}=-(\mathcal
 M^{-1})^{MN}(\pi_N-\mathcal M_{NL}\d_0 D^L),\\
 &   \vddl{\mathcal L_{DP}}{D^M}=\Omega_{MN}(\Delta A_0^N-\d_0 \d^i A_i^N)
    +\d^i(\mathcal M_{MN} \mathcal B^N_i)-\d_0\pi_M-\frac{\d \mathcal
      V}{\d D^M},\\
 &   \vddl{\mathcal L_{DP}}{A^M_0}=-\Omega_{MN}\Delta
    D^N=-\frac{1}{2}\Omega_{MN}\epsilon^{ijk} \d_i\mathcal
    F_{jk}^N,\\
 &   \vddl{\mathcal L_{DP}}{A^M_i}=\Omega_{MN}\d_0
    \mathcal B^{Ni}-\epsilon^{ijk}\d_j(\mathcal M_{MN} \mathcal B^N_k)
    =\half\Omega_{MN}\epsilon^{ijk} \d_0\mathcal F_{jk}^N-\d_j
    (\mathcal M_{MN}\mathcal F^{Nij}).
\end{split}
\end{equation}
The first set of equations then allows one to eliminate the momenta
$\pi_M$ by their own equations of motion. When $\Delta$ is invertible,
the second and third set of equations allow one to solve $D^M$ and
$A_0^M$ by their own equations of motion in the action, which yields
\eqref{eq:symlag}. It is in this sense that these variants of the
double potential formalism are equivalent, but of course not locally
so. The third and fourth set of equations can be written as
\begin{equation}
  \label{eq:55}
 \vddl{\mathcal L_{DP}}{A^M_\mu}=\d_\nu\star G^{\mu\nu}_M,
\end{equation}
when defining
\begin{equation}
\star G^{i0}_{M}=\half\Omega_{MN}\epsilon^{ijk} \mathcal F_{jk}^N,
  \quad \star G_{M}^{ij}=-\mathcal M_{MN}\mathcal F^{Nij}.
\end{equation}

This definition implies that the components of
$G_M=\half G_{Mjk} dx^jdx^k+G_{Mi0}dx^idx^0$ are explicitly given by
\begin{equation}
  \label{eq:57}
  G_{Mjk}=-\Omega_{MN}\mathcal F_{jk}^N,\quad G_{Mi0}=\half
  \epsilon_{ijk}\mathcal M_{MN}\mathcal F^{Njk}.
\end{equation}
After elimination of the $\pi_M$, the action of the theory can
then also be written as the integral of 
$\mathcal{L}_0=\mathcal{L}_{ES}+ \mathcal{L}_V$ with
\begin{equation}
  \mathcal{L}_{ES}=\mathcal L_S-\frac{1}{2}\partial_\mu
  D^M\mathcal{M}_{MN}\partial^\mu D^N-\mathcal V,
  \quad d^4x \mathcal{L}_V=\int_0^1\frac{dt}{t} [G_MF^M][tA^M,D^M,\phi^i].
\end{equation}
so that the scalar sector has been enlarged to $\phi^m=(\phi^i,D^M)$
and the scalar metric and potential are now
$(g_{ij},\mathcal M_{MN})$, respectively $(V,\mathcal V)$. It is thus
a particular case of the actions of the form
\eqref{4.4} studied in section \ref{sec:gauging}.

\subsection{Local BRST cohomology}
\label{sec:local1st}

The master action is given by 
\begin{equation}
  S=\int d^4x\,[ \mathcal L_0+A^{*\mu}_M\partial_\mu C^M]
\end{equation}
with an antifield and ghost sector that is doubled as compared to the
second order covariant formulation.

We then can copy previous results:

(i) $H^g(s)=0$ for $g\leq -3$.

(ii) $H^{-2}(s)$ is doubled: $U^{-2}=\mu^M d^4x C^*_M$ with descent equation

\begin{equation}
s\, d^4x\, C^*_M+d \star A^*_M=0,\quad s \star A^*_M+d
G_M=0,\quad sG_M=0.\label{eq:14bis}
\end{equation}
Characteristic cohomology $H^{n-2}_W(d)$ is then represented by the
2-forms $\mu^MG_M$.

For $g\geq -1$, the discussion in terms of $U$, $W$, $V$ types is the
same as before with indices $I,J,K,\dots \to M,N,O,\dots$ on vector
potentials, ghosts and their antifields, and
$i,j,k\dots\to m,n,o\dots$ on scalar fields.

The obstruction equation for symmetries, equation \eqref{eq:17a},
becomes
\begin{multline}
  \label{eq:17c}
  G_M F^N {f^M}_N+F^MF^Nh_{MN}+dI^{n-1}\\
-(dG_Mg^{M}+[d(g_{ij} \star
  d\phi^j)+\star
  \d_i(\mathcal{L}_{ES}+\mathcal{L}_V)]\Phi^{i}
  +\vddl{\mathcal L_0}{D^M} \Phi^{M})
=0.
\end{multline}

\subsection{Constraints on $W,U$-type cohomology}
\label{sec:absence-w-type}

When there is no explicit $x^\mu$ dependence and $V=0=\mathcal V$,
putting all derivatives of $F^M_{\mu\nu},\phi^i,D^M$ to zero, one
remains with
\begin{equation}
  \label{eq:18c}
  {G_M}|_{{\rm der}=0} F^N{f^M}_N+F^MF^Nh_{MN}
-\star \d_m \mathcal{L}_V
\Phi^{m}|_{{\rm der}=0}=0,
\end{equation}
where ${G_M}|_{{\rm der}=0}$ amounts to replacing $\mathcal
F^M_{\mu\nu}$ by $F^M_{\mu\nu}$ in \eqref{eq:57}. 

Using
$-\d_m\star \mathcal{L}_V=\delta^i_m\half\d_i G_M|_{{\rm der}=0} F^M$,
and the decomposition 
$\Phi^{m}|_{{\rm der}=0}=\Phi^{m}_0+\Phi^{m}_1+\dots$, where the
$\Phi^{m}_{n}$ depend on undifferentiated scalar fields and are
homogeneous of degree $n$ in $F^M_{\mu\nu}$, the
equation implies
\begin{equation}
  \label{eq:26c}
G_M|_{{\rm der}=0}F^N {f^M}_N
+F^MF^Nh_{MN}+\half
  \d_iG_M|_{{\rm der}=0}F^M\Phi^{i}_0=0.
\end{equation}
When taking account that
\begin{equation}
  G_M|_{{\rm der}=0} F^N=d^4x\frac{1}{2} [\Omega_{OM}\epsilon^{ijk}
  F^O_{jk} F^N_{0i}-  \mathcal
  M_{MO} F^O_{jk} F^{Njk}],\label{eq:56}
\end{equation}
this gives an equation of the type
\begin{equation}
  \label{eq:28b}
  \frac{1}{4} d^4x[  \mathcal O_{MN}(\phi)\epsilon^{ijk}
  F^M_{jk} F^N_{0i}-\mathcal P_{MN}(\phi)F^M_{jk}F^{Njk}]=0, 
\end{equation}
where 
\begin{equation}
  \label{eq:29a}
  \mathcal P_{MN}=2\mathcal M_{O(M}{f^O}_{N)}
+ \d_i\mathcal
M_{MN}\Phi^{i}_0, 
\end{equation}
\begin{equation}
  \label{eq:27b}
  \quad  \mathcal
  O_{MN}=2\Omega_{MO}{f^O}_N+2h_{MN},
\end{equation}
and $h_{MN}=h_{NM}$ on account of \eqref{Y1}.  Note that there is one
less term as compared to \eqref{eq:27} since the kinetic term does not
depend on the scalars and also that $\mathcal O_{MN}$ is not
symmetric.

Now both terms have to vanish separately because they involve
different field strengths, 
\begin{equation}
\mathcal P_{MN}=0,\quad \mathcal O_{MN}=0.\label{eq:19a}
\end{equation}
Setting $\phi^i=0=D^M$ then gives
\begin{equation}
\begin{split}
  & f^{(\mathcal M(0))}_{MN}+f^{(\mathcal M(0))}_{NM}
  +(\d_i\mathcal M_{MN})(0)\Phi^{i}_0(0)=0,\\ 
& f^{(\Omega)}_{MN}=-h_{MN}, \label{eq:35a}
\end{split}
\end{equation}
with $f^{(\Omega)}_{MN}=\Omega_{MO}{f^O}_N$. 
Consider first symmetries of $W$-type, i.e., take the case when
the $f$'s vanish. The first equation is then satisfied with
$\Phi^i_0(0)=0$, while the second equation then requires
$h_{MN}$ to vanish. This implies:

{\em There are neither $W$-type symmetries nor $W$-type cohomology in ghost
  numbers $g\geq 0$ for the first order model.}

As a consequence, $h_{MN}={h_u}_{MN}$, and the second of equation \eqref{eq:35a} is
equivalent to 
\begin{equation}
{f_u}^{(\Omega)}_{[MN]}=0,\quad {f_u}^{(\Omega)}_{(MN)}=-{h_u}_{MN}\label{eq:35}.
\end{equation}
It follows that:

{\em The algebra $\mathfrak{g}_U$ is the largest sub-algebra of
  $\mathfrak{sp}(2n_v,\mathbb R)$ that can be turned into symmetries of
  the full theory. All non-trivial $U$-type symmetries require a non-vanishing
  ${h_u}_{MN}$ and thus involve a Chern-Simons term in their Noether
  currents.}

On its own, the first equation of \eqref{eq:35a} is solved for
skew-symmetric $f^{(\mathcal M(0))}_{MN}$ with vanishing
$\Phi^i(0)$. Symmetric $f^{(\mathcal M(0))}_{MN}$ needs a non trivial scalar
symmetry.

For $U$-type cohomologies in higher ghost number $g\geq 0$, the
$k^u_{O_1\dots O_{g+1}}$ tensor has to satisfy \eqref{linf},
which becomes
\begin{equation}
  \label{eq:31}
  {f_u}^{(\Omega)}_{M(N}k^u_{O_1)\dots O_{g+1}}=0.
\end{equation}
The object
$D_{MO_1NO_2\dots O_{g+1}}={f_u}^{(\Omega)}_{MN}k^u_{O_1\dots
  O_{g+1}}$ is then symmetric in the first and third indices because
${f_u}^{(\Omega)}_{MN}$ is symmetric, and antisymmetric in the second
and third indices on account of \eqref{eq:31}. It thus has to vanish,
\begin{multline}
  D_{MO_1NO_2\dots O_{g+1}} =D_{NO_1MO_2\dots O_{g+1}} =-D_{NMO_1O_2\dots
    O_{g+1}}   =-D_{O_1MNO_2\dots O_{g+1}}\\
  =D_{O_1NMO_2\dots
  O_{g+1}} =D_{MNO_1O_2\dots O_{g+1}} =-D_{MO_1NO_2\dots O_{g+1}}.
\end{multline}
It follows that $k^u_{O_1\dots O_{g+1}}=0$:

{\em There are no $U$-type cohomology classes in ghost number $g \geq 0$.}

In particular, there are no $U$-type gaugings even though there are
$U$-type symmetries. We thus recover the results on gaugings of
\cite{Bunster:2010wv} from the current perspective.

\subsection{Remarks on $GL(2n_v)$ transformations}

The two remarks on linear changes of variables from section
\ref{sec:frame-transf} also apply in the first order case. More
precisely, the second remark can be rephrased as follows. 

The general discussion of the structure of the BRST cohomology of the
first order model in sections \ref{sec:local1st} and
\ref{sec:absence-w-type} goes through unchanged for arbitrary
skew-symmetric non-degenerate $\Omega_{MN}$ and symmetric
non-degenerate $\mathcal M_{MN}$. The local BRST cohomology for sets
of $\Omega_{MN}$, $\mathcal M_{MN}$ related by $GL(2n_v, \mathbb R)$
transformations will be isomorphic, whereas explicit results for the
local BRST cohomology do depend on the equivalence classes. For
instance for the symmetries, this is the case when explicitly solving
the obstruction equation \eqref{eq:17c}.
As concerns $\Omega_{MN}$, there is just one equivalence class since
all such matrices are related to a canonical $\Omega_{MN}$, say
$\Omega_{MN}=\delta_{IJ}\epsilon_{ab}$, by a $GL(2n_v, \mathbb{R})$
transformation. Hence, one can restrict oneself to equivalence classes
of $\Omega_{MN}$, $\mathcal M_{MN}$ with canonical $\Omega_{MN}$, and
$\mathcal M_{MN}$'s related by $Sp(2n_v,\mathbb R)$ changes of variables.

The first remark of section \ref{sec:frame-transf} then boils down to
the statement that the algebra $\mathfrak{g}_U$ is the largest
sub-algebra of $\mathfrak{sp}(2n_v,\mathbb R)$ that can be turned into
symmetries of the full theory, in agreement with the discussion of the
previous section. In addition we have recovered there the result that
the gauge algebra remains abelian.

\subsection{Application to the bosonic sector of $\mathcal N=4$
  supergravity}
\label{sec:application1}

For definiteness, let us again concentrate on the bosonic sector of
four dimensional supergravity, without gravity. As in section
\ref{sec:N=4}, we use the standard second order formulation for the
$SL(2,\mathbb R)/SO(2)$ sigma model. Alternatively, one could use a
first order formulation in terms of fields parametrizing
$SL(2,\mathbb R)$, with a first class constraint eliminating the field
for the $SO(2)$ subgroup. It would provide a first order formulation
for all fields and make all global symmetries manifest.

To this scalar action, we first couple one vector field, i.e. add the
action associated to \eqref{eq:49} where $\mathcal V=0$, the indices
$M,N$ take two values $a,b$, $\Omega_{ab}=\epsilon_{ab}$, and
$\mathcal M_{ab}=M^{-1}_{ab}$. The matrix $M$ and its inverse are
given by
\begin{equation}
  \label{eq:2a}
  {M}=\begin{pmatrix} e^\phi & \chi e^\phi \\
\chi e^\phi & \chi^2 e^\phi +e^{-\phi}
\end{pmatrix},\quad 
{{M^{-1}}}=\begin{pmatrix} \chi^2 e^\phi +e^{-\phi} & -\chi e^\phi \\
-\chi e^\phi & e^{\phi}
\end{pmatrix}
\end{equation}
and are such that $M$ transforms as $M\to g^TMg $ under an
$SL(2,\mathbb R)$ transformation. The model is invariant under
$SL(2,\mathbb R)$ if the other fields transform as $A^a\to (g^T A)^a$,
$D^a\to (g^T D)^a$, $\pi_a\to (g^{-1}\pi)_a$ because $SL(2,\mathbb R)$
transformations are symplectic, $g\epsilon g^T=\epsilon$.

For the $U$-type symmetries, equation \eqref{eq:35} requires
$f^{(\epsilon)}_{ab}=\epsilon_{ac} {f^c}_b$ to be symmetric, so there
are at most 3 linearly independent solutions. According to the above
discussion, all of these give rise to symmetries, which need
${h_u}_{ab}$ and also $\Phi^i_u$. The $U$-type symmetries constitute
the $\mathfrak{sl}(2,\mathbb R)$
electric symmetry algebra. 

We now consider the coupling to six vector fields in the different
formulations of section \ref{sec:N=4}.  For the $SO(6)$ invariant
model, $M=(I,a)$, $\Omega_{MN}=\delta_{IJ}\epsilon_{ab}$ and
$\mathcal M_{MN}= \delta_{IJ}{M^{-1}}_{ab}$, while the dual
formulation corresponds to $\mathcal M_{MN}=
\delta_{IJ}M_{ab}$. Finally, in the $SO(3)\times SO(3)$ formulation
$\mathcal M_{MN}=(M^{-1}_{ab}\delta_{AB},M_{ab}\delta_{A'B'})$.

It then follows from \eqref{eq:35a} that both in the $SO(6)$ invariant
formulation and in the dual formulation, the electric symmetry algebra
is $\mathfrak{sl}(2,\mathbb R)\oplus \mathfrak{so}(6)$, where the
$\mathfrak{sl}(2,\mathbb R)$ transformations on the vectors $A^{(I,a)}$
and on $D^{(I,a)},\pi_{(J,b)}$ in the dual formulation corresponds to
the infinitesimal version of the above transformations where
$g^T\to g^{-1}$.

Finally, in the $SO(3)\times SO(3)$ formulation, the electric symmetry
algebra is also $\mathfrak{sl}(2,\mathbb
R)\oplus\mathfrak{so}(6)$. This is so because the $SL(2,\mathbb R)$
element $\epsilon$ is such that $M=\epsilon^TM^{-1} \epsilon$.

\section{Conclusions and comments}
\label{sec:conclusions}

In this work we have systematically analyzed gaugings of vector-scalar
models through a standard deformation theoretic approach. In the case
of gauge systems, this is most naturally done in the BV-BRST antifield
formalism. We have shown that different types of symmetries behave
differently when one tries to gauge them. The method allows one to
find all the infinitesimal gaugings and higher order cohomology
classes once all symmetries are known.

The symmetries are classified into $U$, $W$ and $V$-types. Only
$U$-type symmetries give rise to gaugings that deform the abelian
gauge algebra. They contain the standard ``Yang-Mills''
deformations. The $W$-type symmetries contain the topological gaugings
of \cite{deWit:1987ph}. The Noether currents of both these types of
symmetries are the only ones that cannot be redefined so as to be
gauge invariant.  We have treated explicit examples, for which all
symmetries of $U$, $W$-types have been computed (in the $x^\mu$
independent case considered here).

For the models explicitly considered in the article, we have found that the only possible gaugings
of $U$ and $W$-types are the ones previously considered in the literature, namely
Yang-Mills and topological couplings  among the
gauge fields, with minimal couplings of the scalars.

In order to achieve complete results, one should also compute the
$V$-type symmetries which admit gauge invariant Noether currents. This
is very much a model-dependent question that requires the use of more
standard symmetry techniques (see e.g.,
\cite{Anco2003,Pohjanpelto:2008st}). However, we have shown in Section
\ref{sec:AntiMap} that given the graded structure of the antibracket
map, the leading obstruction to extend first order deformations of
$U$-type to second order, leading to the Jacobi identity for the
structure constants, cannot be eliminated by adding $V$-terms.

Furthermore, in some cases, for instance when one imposes Poincar\'e
invariance as relevant to relativistic theories, the $V$-type
symmetries can be shown to be absent \cite{Torre:1995kb}. It turns out
that the effect of coupling the models to Einstein gravity justifies
this assumption \cite{Barnich:1995ap} and simplifies the problem. It
would be interesting to see if it also justifies the ansatz for the
electric symmetry algebra.

We have analyzed the problem in the second order Lagrangian and in the
first order manifestly duality invariant formulation, both of which
are non-locally related in space (but not in time). The results are
very different: whereas the former formulation allows for standard
gaugings, the latter formulation allows for more (generalized)
symmetries of $U$-type, but none of those can be gauged. This is
because the analysis is performed in each formulation by insisting on
space-time locality. To go beyond such no-go results, one should
presumably try to work in a controlled way with deformations that are
spatially non-local.

\subsection*{Acknowledgments}

\addcontentsline{toc}{section}{Acknowledgments}

GB, NB and MH thank the Laboratoire de Physique Th\'eorique de l'Ecole
Normale Sup\'erieure for kind hospitality. BJ thanks ULB for its
generous hospitality and Mario Trigiante for information.
GB is grateful to F.~Brandt for earlier
collaborations on this subject. This work was partially supported by
the ERC Advanced Grant ``High-Spin-Grav", by FNRS-Belgium (convention
FRFC PDR T.1025.14 and convention IISN 4.4503.15) and by the
``Communaut\'e Fran\c{c}aise de Belgique" through the ARC program.  NB
is Senior Research Associate of the F.R.S.-FNRS. VL is Research
Fellow of the F.R.S.-FNRS.

\begin{appendix}

\section{Conventions and notation}
\label{sec:notations}

The components of the Minkowski metric are given, in inertial 
coordinates in which we work, by the mostly plus expression 
$\eta_{\mu\nu}={\rm diag}(-1,+1,\dots,+1)\,$. 
The symbol $\epsilon_{\mu_1\dots\mu_n}$ denotes the completely 
antisymmetric Levi-Civita density with the convention that 
$\epsilon^{01\dots n-1}=1\,$ so that $\epsilon_{01\dots n-1}=-1\,$. 
A local basis of anticommuting exterior differential 1-forms 
is given by the family $(dx^\mu)_{\mu = 0, \ldots, n-1\,}$. 
The wedge product symbol $\wedge$ will always be omitted.   

We will sometimes use the notation 
$(d^{ n - p }x)_{\mu_{n-p+1}\dots\mu_n} := -\frac{1}{p!(n -
  p)!}dx^{\mu_{1}}\dots dx^{\mu_{n-p}}\epsilon_{\mu_1\dots\mu_n}$
for $1\leqslant p \leqslant n\,$, and $d^nx:=dx^0\dots dx^{n-1}\,$.
The Hodge dual of a differential $p\,$-form
$\omega^p \equiv \frac{1}{p!}\,dx^{\mu_1}\dots
dx^{\mu_p}\omega_{\mu_1\dots\mu_p}$,
is the $n-p\,$-form given, in our convention, by 
\begin{align}
\star\,\omega^p&=\tfrac{1}{p!(n-p)!}\,dx^{\nu_1}\dots
dx^{\nu_{n-p}}\epsilon_{\nu_1\dots\nu_{n-p}\mu_{n-p+1}\ldots\mu_n}
\omega^{\mu_{n-p+1}\dots\mu_n}=-
(d^{ n - p }x)_{\mu_{n-p+1}\dots\mu_n}
\omega^{\mu_{n-p+1}\dots\mu_n}\;. 
\nonumber
\end{align}
As a consequence, the exterior differential of the dual of a 
$p\,$-form reads 
\begin{align}
d \star \omega^p =-(-)^{n-p} (d^{n-p+1}x)_{\nu_1\dots\nu_{p-1}}
\,\partial_{\mu}\omega^{\mu\nu_1\dots\nu_{p-1}}\;.
\end{align}

\section{Antibracket maps and descents}
\label{sec:antibr-maps-desc}

As discussed in section \ref{sec:BRST}, the first
obstruction to extending infinitesimal deformations to finite ones is
controlled by the antibracket map. We show here how the antibracket
map behaves with respect to the length of shortest non trivial
descent, i.e., the ``depth''.

Since covariantizable and non-covariantizable currents as elements of
$H^{-1,n}(s|d)$, and the associated infinitesimal deformations as
elements of $H^{0,n}(s|d)$ are distinguished by the property that the
depth is $1$, respectively deeper than one, the following will be
relevant when studying the obstruction to infinitesimal deformations.

{\bf Proposition:}

{\em The depth of an image of the antibracket map is less or equal to the depth of
  its most shallow argument.}

{\bf Proof:} 

Consider
$[\omega^{g_1,n}_{l_1}],[\omega^{g_2,n}_{l_2}]\in H^{*,n}(s|d)$, where
we can assume without loss of generality that $l_1\geqslant l_2$. For the
antibracket, let us not choose the expression with Euler-Lagrange
derivatives on the left and right that is graded antisymmetric without
boundary terms, but rather the one that satisfies a graded Leibniz
rule on the right
\begin{equation}
  \label{eq:B2}
  (\omega^{g,n},\cdot)_{\rm
    alt}=\d_{(\nu)}\vddr{(-\star\omega^{g,n})}{\phi^A}\ddll{\cdot}{\d_{(\nu)}\phi^*_A}
    -(\phi^A\leftrightarrow \phi^*_A), 
\end{equation}
and the following version of the graded Jacobi identity without
boundary terms,
\begin{equation}
  \label{eq:B3}
  (\omega^{g_1,n},(\omega^{g_2,n},\cdot)_{\rm alt})_{\rm
    alt}=((\omega^{g_1,n},\omega^{g_2,n})_{\rm alt},\cdot)_{\rm
    alt}+(-)^{(g_1+1)(g_2+1)}(\omega^{g_2,n},(\omega^{g_1,n},\cdot)_{\rm alt})_{\rm
    alt}
\end{equation}
(see appendix B of
\cite{Barnich:1996mr} for details and a proof). 
Furthermore, 
\begin{equation}
  (\omega^{g,n},d (\cdot))_{\rm
    alt}=(-)^{g+1}d((\omega^{g,n}, \cdot)_{\rm
    alt}),\quad (d\omega^{g+1,n-1},\cdot)_{\rm alt}=0. \label{eq:B5}
\end{equation}
Let $S=\int (-\star \cL)$ be the BV master action. We have
$s\cdot=(-\star \cL,\cdot)_{\rm
  alt}$. 
Using these properties, we get 
  \begin{equation}
    \label{eq:B4}
    s(\omega^{g_1,n}_{l_1},\omega^{g_2,n}_{l_2})_{\rm
      alt}+d((\omega^{g_1,n}_{l_1},\omega^{g_2+1,n-1}_{l_2})_{\rm
      alt})=0,
\dots, s (\omega^{g_1,n}_{l_1},\omega^{g_2+l_2,n-l_2}_{l_2})_{\rm
      alt}=0, 
  \end{equation}
which proves the proposition. \qed

By using $[(\omega^{-1,n},\omega^{g,n})]\in H^{g,n}(s|d)$, it follows
that :

(i) Characteristic cohomology in degree $n-1$ described by
$H^{-1,n}(s|d)$ is a Lie algebra. It is the Lie algebra of non trivial
global symmetries. It also describes the Dirac or Dickey bracket
algebra of non trivial conserved currents (up to constants or more
generally topological classes),

(ii) $H^{-2,n}(s|d)$ is a module thereof (module structure of
flux charges - Gauss or ADM type surface charges - under global symmetries). The proposition
gives rise for instance to the following refinements:

{\bf Corollary:} Covariantizable characteristic cohomology in form
degree $n-1$ forms an ideal in the Lie algebra of characteristic
cohomology in form degree $n-1$. The module action of covariantizable
characteristic cohomology of degree $n-1$ on characteristic cohomology
in degree $n-2$ is trivial. 

Similar results hold for the associated infinitesimal deformations.

\section{Derivation of Equation \eqref{eq:Utrasf}}
\label{app:derivation}

In this appendix, we derive formula \eqref{eq:Utrasf} for the variation $\delta_u G_I = - (U_u, G_I)$ of the two-form $G_I$ under a $U$-type symmetry. This is done in two steps:
\begin{enumerate}
\item First, we show that
\begin{equation}\label{eq:step1}
\delta_u G_I + (f_u)\indices{^J_I} G_J \approx c_{IJ} F^J + d(\text{invariant})
\end{equation}
for some constants $c_{IJ}$.
\item Then, we prove that the $c_{IJ}$ take the form
\begin{equation}\label{eq:step2}
c_{IJ} = - 2 (h_u)_{IJ} + \lambda^w_u (h_w)_{IJ},
\end{equation}
where the constants $(h_u)_{IJ}$ and $(h_w)_{IJ}$ are those appearing in the currents associated with $U_u$ and $W_w$ respectively.
\end{enumerate}
The proof is given in the case where the Lagrangian (or, equivalently, $G_I$) does not depend on the derivatives of $F^I_{\mu\nu}$.

\subsubsection*{A lemma}

The proof of the above
 steps uses the following result on the $W$-type cohomology classes (with $g=-1$):
\begin{equation}\label{eq:Wlemma}
t_{IJ} F^I F^J \approx d(\text{invariant}) \;\Rightarrow\; t_{IJ} = \sum_{w} \lambda^w (h_w)_{IJ} \,\text{ for some } \lambda^w .
\end{equation}
This is proven as follows: $t_{IJ} F^I F^J \approx d(\text{invariant})$ implies that
\begin{equation}\label{eq:appt}
t_{IJ} F^I F^J + dI + \delta k = 0
\end{equation}
for some gauge invariant $I$ and some $k$ of antifield number $1$, where $\delta$ is here the Koszul-Tate differential. Now, it is proven in \cite{Barnich:2000zw} that $k$ must be gauge invariant; hence, it can be written as
\begin{equation}\label{eq:appk}
k = \hat{K} + d R, \quad \hat{K} = d^4x [ A^{*\mu}_I g^I_\mu + \phi^*_i \Phi^i ]
\end{equation}
for some gauge invariant $R$, $g^I_\mu$ and $\Phi^i$. Indeed, derivatives acting on the antifields contained in $k$ are pushed to the term $dR$ by integration by parts, leaving the form \eqref{eq:appk} where $\hat{K}$ contains only the undifferentiated antifields. Putting this back in \eqref{eq:appt} and using the fact that $\delta \hat{K} = s \hat{K}$ because $\hat{K}$ is gauge invariant, we get
\begin{equation}
s \hat{K} + d\left( t_{IJ} A^I F^J + J \right) = 0
\end{equation}
for some gauge invariant $J = I - \delta R$. This shows that $\hat{K}$ is a $W$-type cohomology class: we can therefore expand $\hat{K}$ in the $W_w$ basis as $\hat{K} = \sum \lambda^w W_w$. In particular, this implies that $t_{IJ} = \sum \lambda^w (h_w)_{IJ}$, which proves the lemma.

\subsubsection*{First step}

We start from the chain of descent equations involving $G_I$,
\begin{equation}\label{eq:descCstar}
s\, d^4x\, C^*_I+d \star A^*_I=0,\quad s \star A^*_I+d
G_I=0,\quad sG_I=0.
\end{equation}
Applying $(U_u, \cdot)_\text{alt}$ to this chain, we get
\begin{align}
s \left[ \,d^4x\, (f_{u})\indices{^J_I} C^*_J \right] + d \left[ (f_{u})\indices{^J_I} \star A^*_J +
\frac{\delta K_u}{\delta A^I} \right] &= 0, \\
s \left[ (f_{u})\indices{^J_I} \star A^*_J +
\frac{\delta K_u}{\delta A^I} \right] + d \left[ - \delta_u G_I \right] &= 0, \\
s \left[ - \delta_u G_I \right] &= 0,
\end{align}
which can be simplified to
\begin{align}
d \left( \frac{\delta K_u}{\delta A^I} \right) &= 0, \label{eq:dK}\\
s \left( \frac{\delta K_u}{\delta A^I} \right) + d \left( - \delta_u G_I - (f_{u})\indices{^J_I} G_J \right) &= 0, \label{eq:sK}\\
s \left( - \delta_u G_I \right) &= 0,
\end{align}
using equations \eqref{eq:descCstar} again.
Equation \eqref{eq:dK} implies that
\begin{equation}
\frac{\delta K_u}{\delta A^I} = d \eta^{-1,2}
\end{equation}
for some $\eta^{-1,2}$ of ghost number $-1$ and form degree $2$. Because the left-hand side is gauge invariant and $\eta^{-1,2}$ is of form degree two, $\eta^{-1,2}$ must also be gauge invariant. This follows from theorems on the invariant cohomology of $d$ in form degree $2$ \cite{Brandt:1990gy,Dubois-Violette:1992ye}. Equation \eqref{eq:sK} implies then
\begin{equation}
d\left( \delta_u G_I + (f_{u})\indices{^J_I} G_J + s\eta^{-1,2} \right) = 0 ,
\end{equation}
i.e.
\begin{equation}\label{eq:dGdeta}
\delta_u G_I + (f_{u})\indices{^J_I} G_J + s\eta^{-1,2} = d \eta^{0,1}
\end{equation}
for some $\eta^{0,1}$ of ghost number $0$ and form degree $1$. Again, the left-hand side of this equation is gauge invariant: results on the invariant cohomology of $d$ in form degree $1$ \cite{Brandt:1990gy,Dubois-Violette:1992ye} now imply that the non-gauge invariant part of $\eta^{0,1}$ can only be a linear combination of the one-forms $A^I$,
\begin{equation}
\eta^{0,1} = c_{IJ} A^J + \text{(gauge invariant)} .
\end{equation}
Plugging this back in equation \eqref{eq:dGdeta} and using the fact that $s\eta^{-1,2} \approx 0$ (since $\eta^{-1,2}$ is gauge invariant), we recover equation \eqref{eq:step1}. This concludes the first step of the proof.

\subsubsection*{Second step}

For the second step, we introduce
\begin{equation}
N = - \int \!d^4x\,( C^*_I C^I + A^{*\mu}_I A^I_\mu), \quad \hat{N} = (N, \cdot)_\text{alt} .
\end{equation}
The operator $\hat{N}$ counts the number of $A^I$'s and $C^I$'s minus the number of $A^*_I$'s and $C^*_I$'s. Because it carries ghost number $-1$, it commutes with the exterior derivative, $\hat{N} d = d \hat{N}$.
Applying this operator to the equation
\begin{equation}
s U_u + d \left[ (f_u)\indices{^I_J} (\star A^*_I C^J + G_I A^J) + (h_u)_{IJ} F^I A^J  + J_u \right] = 0
\end{equation}
gives
\begin{equation}\label{eq:NsU}
(\int\! G_I F^I, U_u)_\text{alt} + d\left[ (f_u)\indices{^I_J} (\hat{N} + 1)(G_I) A^J + 2 (h_u)_{IJ} F^I A^J  + \hat{N}(J_u) \right] \approx 0 .
\end{equation}
The second term is evident. The first term is
\begin{align}
\hat{N}(sU_u) = (N, (S,U_u)_\text{alt})_\text{alt} &= ( (N, S) , U_u)_\text{alt} + (S, (N,U_u)_\text{alt})_\text{alt}
\end{align}
according to the graded Jacobi identity.
The counting operator $\hat{N}$ kills the $A^{*\mu}_I \partial_\mu C^I$ term in the master action $S$, which implies
\begin{equation}
(N, S) = \int\!d^4x\, A^I_\mu \frac{\delta \mathcal{L}_V}{\delta A^I_\mu} = \int\!d^4x\, A^I_\mu \partial_\nu(\star G_I)^{\mu\nu} = \int\! G_I F^I .
\end{equation}
Similarly, $\hat{N}$ kills the first two terms of $U_u$, leaving $\hat{N} U_u = \hat{N} K_u$ which is gauge invariant. This implies $(S, (N,U_u)_\text{alt})_\text{alt} = s(\hat{N} U_u) \approx 0$.
Therefore, we have indeed
\begin{equation}
\hat{N}(sU_u) \approx (\int G_I F^I, U_u)_\text{alt}
\end{equation}
which proves equation \eqref{eq:NsU}.

We now compute $(\int G_I F^I, U_u)_\text{alt}$ using the result of the first step. We have
\begin{equation}
(\int G_I F^I, U_u)_\text{alt} = \frac{\delta (G_K F^K)}{\delta A^I_\mu} \, \delta_u A^I_\mu + \frac{\delta (G_K F^K)}{\delta \phi^i} \, \delta_u \phi^i .
\end{equation}
This looks like the $U$-variation $\delta_u (G_I F^I)$, but it is not because there are Euler-Lagrange derivatives. For a top form $\omega$, the general rule is \cite{Andersonbook}
\begin{equation}
\delta_Q \omega = Q^a \frac{\delta \omega}{\delta z^a} + d \rho, \quad \rho = \partial_{(\nu)} \left[ Q^a \frac{\delta}{\delta z^a_{(\nu)\rho}} \frac{\partial \omega}{\partial dx^\rho} \right] .
\end{equation}
In our case, this becomes
\begin{align}
\delta_u (G_I F^I) &= \frac{\delta (G_K F^K)}{\delta A^I_\mu} \, \delta_u A^I_\mu + \frac{\delta (G_K F^K)}{\delta \phi^i} \, \delta_u \phi^i + d\rho_A + d\text{(inv)} ,\\
\rho_A &= \partial_{(\nu)} \left( (f_u)\indices{^I_J} A^J_\mu \frac{\delta}{\delta A^I_{\mu, (\nu)\rho}} \frac{\partial (G_K F^K)}{\partial dx^\rho} \right) .
\end{align}
Using property \eqref{eq:step1} and putting together the terms of the form $d\text{(invariant)}$, we get then from \eqref{eq:NsU}
\begin{equation}
(c_{IJ} + 2 (h_u)_{IJ} ) F^I F^J + d \left[ (f_u)\indices{^I_J} A^J (\hat{N} + 1)(G_I) - \rho_A \right] + d \text{(inv)} \approx 0 .
\end{equation}
Now, it is sufficient to prove that
\begin{equation}\label{eq:dxy}
 d \left[ (f_u)\indices{^I_J} A^J (\hat{N} + 1)(G_I) - \rho_A \right] \approx d \text{(inv)}.
\end{equation}
Indeed, this implies $(c_{IJ} + 2 (h_u)_{IJ} ) F^I F^J \approx d \text{(inv)}$, which in turn gives
\begin{equation}
c_{IJ} = - 2 (h_u)_{IJ} + \lambda^w_u (h_w)_{IJ}
\end{equation}
for some constants $\lambda^w_u$ using property \eqref{eq:Wlemma} of the $W$-type cohomology classes.

\subsubsection*{Proof of \eqref{eq:dxy}}

We will actually prove the stronger equation
\begin{equation}\label{eq:xy}
\rho_A = (f_u)\indices{^I_J} A^J (\hat{N} + 1)(G_I)
\end{equation}
in the case where $G_I$ depends on $F$ but not on its derivatives.

To do this, we can assume that $G_I$ a homogeneous function of degree $n$ in $A^I$, i.e. $\hat{N}(G_I) = n G_I$. If it is not, we can separate it into a sum of homogenous parts; the result then still holds because equation \eqref{eq:xy} is linear in $G_I$.

In components, equation \eqref{eq:xy} is
\begin{equation}
\frac{1}{2}\partial_{(\nu)} \left( (f_u)\indices{^I_J} A^J_\mu \frac{\delta}{\delta A^I_{\mu, (\nu)\rho}} G_{K\sigma\tau} F^K_{\lambda\gamma} \varepsilon^{\sigma\tau\lambda\gamma} \right) = (n+1) (f_u)\indices{^I_J} A^J_\lambda G_{I\sigma\tau} \varepsilon^{\rho\lambda\sigma\tau} .
\end{equation}
Under the homogeneity assumption $\hat{N}(G_I) = n G_I$, we have
\begin{equation}
G_{K\sigma\tau} F^K_{\lambda\gamma} \varepsilon^{\sigma\tau\lambda\gamma} = 4 (n+1) \mathcal{L}_V .
\end{equation}
Equation \eqref{eq:xy} now becomes
\begin{equation}\label{eq:xyL}
\frac{1}{2}\partial_{(\nu)} \left( (f_u)\indices{^I_J} A^J_\mu \frac{\delta \mathcal{L}_V}{\delta A^I_{\mu, (\nu)\rho}} \right) = \frac{1}{4} (f_u)\indices{^I_J} A^J_\lambda G_{I\sigma\tau} \varepsilon^{\rho\lambda\sigma\tau} .
\end{equation}
We now use the fact that $G_I$ does not depend on derivatives of $F$, which implies that the higher order derivatives $\partial_{(\nu)}$ are not present and that the Euler-Lagrange derivatives are only partial derivatives. We then have
\begin{equation}
\frac{1}{2} \frac{\delta \mathcal{L}_V}{\delta A^I_{\mu,\rho}} = \frac{\delta \mathcal{L}_V}{\delta F^I_{\rho\mu}} = \frac{1}{4} \varepsilon^{\rho\mu\sigma\tau} G_{I\sigma\tau}
\end{equation}
(see \eqref{eq:47}), which proves \eqref{eq:xyL} in this case.

\end{appendix}

\addcontentsline{toc}{section}{References}

\providecommand{\href}[2]{#2}\begingroup\raggedright\endgroup

\end{document}